\documentclass[twocolumn,superscriptaddress]{revtex4}
\usepackage{amsfonts,amsmath,amssymb}
\usepackage[pdftex]{graphicx}
\usepackage{algorithm}
\usepackage{algorithmic}

\renewcommand{\geq}{\geqslant}

\begin{document}

\title{Sustaining the Internet with hyperbolic mapping}

\author{Mari{\'a}n Bogu{\~n}{\'a}}

\affiliation{Departament de F{\'\i}sica Fonamental, Universitat de Barcelona, Mart\'{\i} i Franqu\`es 1, 08028 Barcelona, Spain}

\author{Fragkiskos Papadopoulos}

\affiliation{Department of Electrical and Computer Engineering, University of Cyprus, Kallipoleos 75, Nicosia 1678, Cyprus}

\author{Dmitri Krioukov}
\affiliation{Cooperative Association for Internet Data Analysis (CAIDA), University of California, San Diego (UCSD), La Jolla, CA 92093, USA}

\begin{abstract}
The Internet infrastructure is severely stressed. Rapidly growing
overheads associated with the primary function of the
Internet---routing information packets between any two computers in
the world---cause concerns among Internet experts that the existing
Internet routing architecture may not sustain even another decade.
Here we present a method to map the Internet to a hyperbolic space.
Guided with the constructed map, which we release with this paper,
Internet routing exhibits scaling properties close to theoretically
best possible, thus resolving serious scaling limitations that the
Internet faces today. Besides this immediate practical viability,
our network mapping method can provide a different perspective on
the community structure in complex networks.
\end{abstract}

\maketitle

\section{Introduction}

In the Information Age, the Internet is becoming a {\it de facto\/}
public good, akin to roads, airports, or any other critical
infrastructure~\cite{Gehring04-book}. More than a billion people are
estimated to use the Internet every day to communicate, search for
information, share data, or do business~\cite{internetusage}. On-line
social networks are becoming an integral part of human social
activities, increasingly affecting human psychology~\cite{LaPe09}.
Underlying all these processes is the Internet infrastructure,
composed, at the large scale, of connections between Autonomous
Systems (ASs). An AS is, roughly, a part of the Internet owned and
administered by the same organisation~\cite{HaBa96}. ASs range in
size from small companies, or even private users, to huge
international corporations. There is no central Internet authority
dictating to any AS what other ASs to connect to. Connections
between ASs are results of local independent decisions based on
business agreements between AS pairs. This lack of centralised
engineering control makes the Internet a truly self-organised
system, and poses many scientific challenges. The one we address
here is the sustainability of Internet growth.

The Internet has been growing fast according to all
measures~\cite{DhDo08-short,Carpenter09}. For example, the number of ASs
increases by approximately $2,400$ every year~\cite{DhDo08-short}. Despite
its growth, the Internet must sustainably perform its primary
task---routing information packets between any two computers in the
world. But can this function be really sustained? To route
information to a given destination in the Internet today, all ASs
must collectively discover the best path to each possible
destination, based on the current state of the global Internet
topology. As the number of destinations grows quickly, the amount of
information each AS has to maintain becomes a serious scalability
concern, endangering the performance and stability of the
Internet~\cite{iab-raws-report-phys}. Worse yet, the Internet is not
static. Its topology changes constantly due to failures of existing
links and nodes, or appearances of new ones. Each time such a change
occurs anywhere in the Internet, the information about this event
must be diffused to all ASs, which have to quickly process it to
recompute new best routes. The constantly increasing size and
dynamics of the Internet thus leads to immense and quickly growing
routing overheads, causing concerns among Internet experts that the
existing Internet routing architecture may not sustain even another
decade~\cite{iab-raws-report-phys,AtBe09,GoGa09,Gammon10}; parts of
the Internet have started sinking into black holes
already~\cite{KaBa08}.

The scaling limitations with existing Internet routing stem from the
requirement to have a current state of the Internet topology
distributed globally. Such global knowledge is unavoidable since
routing has no source of information other than the network
topology. Routing in these conditions is equivalent to routing using
a hypothetical road atlas, which has no geographic information, but
just lists road network links, which are pairs of connected road
intersections, abstractly identified. This analogy with road routing
suggests that there are better ways to find paths in networks.
Suppose we want to travel from one geographic place to another.
Given the geographic coordinates of our starting point and
destination, we can readily tell what direction brings us closer to
our destination. We see that a coordinate system in a geometric
space, coupled with a representation of the world in this space,
simplify drastically our routing task. For simple and efficient
network routing we thus need a map.
Constructing such a map for the Internet boils down to assigning to
each AS its coordinates in some geometric space, and then using this
space to forward information packets in the right directions toward
their destinations. {\em Greedy forwarding\/} implements this
routing in the right direction: upon reading the destination address
in the packet, the current packet holder forwards the packet to its
neighbour closest to the destination in the space.
This greedy strategy to reach a destination is efficient only if the
network map is congruent with the network topology. In the analogy
with road routing, for example, this congruency condition means that
there should exists a road path that stays approximately close to
the geographic geodesic between the trip's starting and ending
points. If the congruency condition holds, then the advantage of
greedy forwarding is twofold. First, the only information that ASs
must maintain is the coordinates of their neighbours. That is, ASs do
not have to keep any per-destination information. Second, once ASs
are given their coordinates, these coordinates do not change upon
topological changes of the Internet. Therefore, ASs do not have to
exchange any information about ever-changing Internet topology.
Taken together, these two improvements essentially eliminate the two
scaling limitations mentioned above.

In our recent work~\cite{BoKrKc08,BoKr09,KrPaBo09,PaKr10} we have
shown that greedy forwarding is indeed efficient in
Internet-like {\em synthetic\/} networks embedded in geometric
spaces, and that this efficiency is maximised if the space is
hyperbolic. However, putting these ideas in practice needs a crucial
piece of information: a map of the {\em real\/} Internet in a
hyperbolic space. Here we present a method to find such a map.

Our method uses statistical inference techniques to find coordinates for
each AS in the hyperbolic space underlying the Internet. Guided by
the inferred coordinates, greedy forwarding in the Internet achieves
efficiency and robustness, similar to those in synthetic networks.
We also find that the method maps geo-politically close ASs close to
each other in the hyperbolic space. This finding suggests that our
mapping method can be used for soft community detection in real
networks, where by {\it soft communities\/} we mean groups of
geometrically close nodes.

\section{The model}

To build a geographic map, one first has to model the Earth surface,
e.g., by assuming that it is a sphere. Similarly, we also need a geometric model of the Internet space to
build our map. The simplest candidate space is also a sphere, or
even a circle, on which nodes are uniformly distributed, and
connected by an edge with probability $p(d)$ decreasing as a
function of distance $d$ between nodes, conceptually similar to
random geometric graphs~\cite{Penrose03-book}. However, this model
fails to capture basic properties of the Internet topology,
including its scale-free node degree distribution.
In~\cite{SeKrBo08}, we showed that to generate realistic network
topologies in this geometric approach, we first have to assign to
nodes their expected degrees $\kappa$ drawn from a power-law
distribution, and then connect pairs of nodes with expected degrees
$\kappa$ and $\kappa'$ with probability $p(\chi)$, where $\chi$ is
distance $d$ rescaled by the product of the expected degrees, $\chi
\sim d/(\kappa \kappa')$. We thus have a hybrid model that mixes
geometry and topology---geometric characteristics, distances $d$
used in random geometric graphs, come in tandem with topological
characteristics, expected degrees $\kappa$ used in classical
configuration models of random power-law graphs~\cite{ChLu02b}. If
we associate the expected degree $\kappa$ of a node with its mass,
then the connection probability $p(d/(\kappa \kappa'))$, which is a
measure of the interaction strength between two nodes, resembles
Newton's law of gravitation. Therefore we call this model {\em
Newtonian.} However, according to Einstein, we can treat gravity in
purely geometric terms if we accept that the space is no longer
flat, i.e., if it is non-Euclidean. Following this philosophy we
showed in~\cite{KrPa09,KrPa10} that the Newtonian model is isomorphic to a
purely geometric network model with node degrees transformed into a
geometric coordinate making the space hyperbolic, i.e., negatively
curved. We call this model {\em Einsteinian.}

\begin{figure}
\centerline{\includegraphics[width=3.5in]{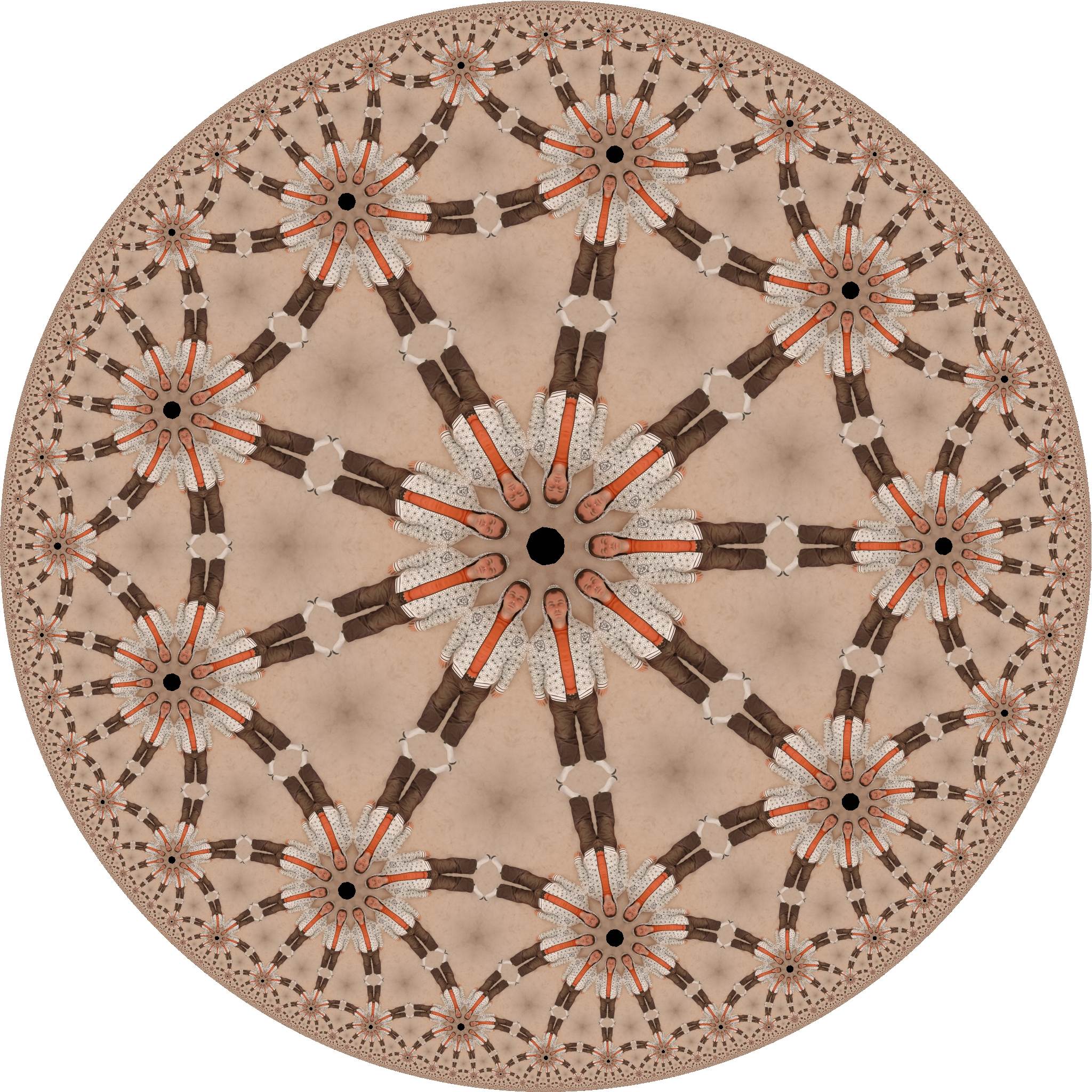}}
\caption{The exponentially growing number of people lying on the
hyperbolic floor illustrates the exponential expansion of hyperbolic
space. All people are of the same hyperbolic size. The {\tt
Poincar\'e\/} tool developed by Bill Horn is used to construct the
tessellation of the hyperbolic plane in the Poincar\'e disc model
with the Schl\"afli symbol $\{9,3\}$, rendering an image of the last
author. \label{fig:0}}
\end{figure}

\begin{figure}
\centerline{\includegraphics[width=3.5in]{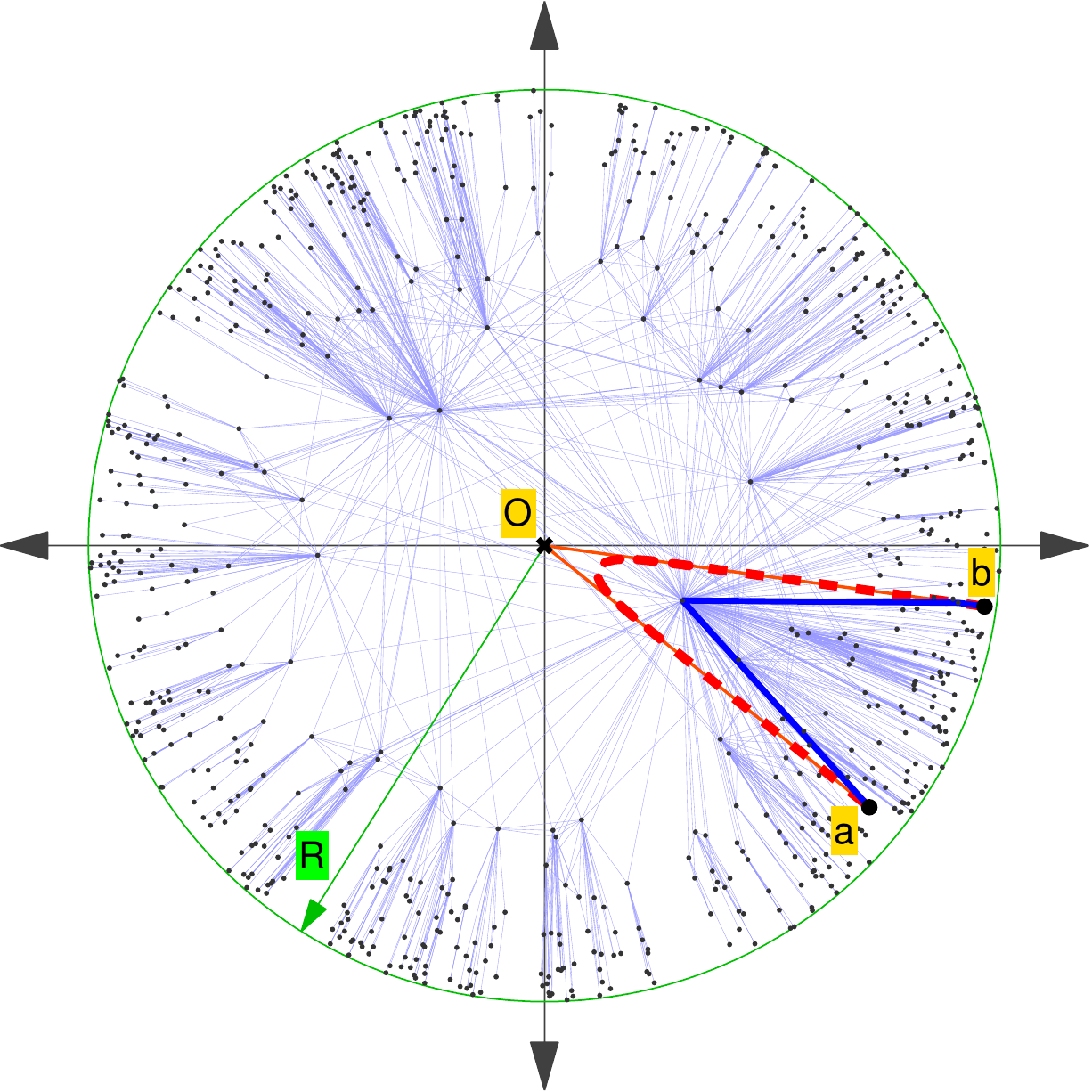}}
\caption{Connection between hyperbolic geometry and scale-free
topology of complex networks illustrated by a synthetic network in
the Einsteinian model. All nodes lie within a hyperbolic disc of
radius $R$. The radial density of nodes grows exponentially with the
distance from the origin $O$, while their average degree
exponentially decreases, yielding a scale-free degree distribution.
The red lines show triangle $Oab$ made of the hyperbolic geodesics
connecting origin $O$ and two nodes $a$ and $b$. Geodesics
$\overline{Oa}$ and $\overline{Ob}$ are the solid lines, while
geodesic $\overline{ab}$ is the dashed curve. The thick blue links
show the shortest path between nodes $a$ and $b$ in the
network.\label{fig:1}}
\end{figure}

The main property of hyperbolic geometry is the exponential
expansion of space illustrated in Fig.~\ref{fig:0}. For example, the
area $A(r)$ of a two-dimensional hyperbolic disc of radius $r$ grows
with $r$ as $A(r) \sim e^r$. Consequently, if we distribute nodes
uniformly or quasi-uniformly over a hyperbolic disc, then from the Euclidean
perspective their density will grow exponentially with the distance
from the disc centre. We illustrate this effect in Fig.~\ref{fig:1},
where we visualise a small-size sample network generated by our
Einsteinian model. In the model, nodes are indeed distributed
(quasi-)uniformly within a hyperbolic disc of radius $R$, which is a
function of the network size. We see that the angular node density
appears uniform, but the radial one does not---the number of nodes
grows exponentially as we move away from the origin. The figure also
shows a triangle connecting origin $O$, and two nodes $a$ and $b$ by
hyperbolic geodesics, i.e., hyperbolically straight lines. The two
geodesics emanating from the origin $O$, $\overline{Oa}$ and
$\overline{Ob}$, are radial straight lines, and their hyperbolic
lengths $x$ are equal to the radial coordinates of $a$ and $b$:
$x_{Oa}=r_a$ and $x_{Ob}=r_b$. However, the hyperbolic geodesic
between nodes $a$ and $b$ does not appear as a Euclidean straight
line, and its length is given by the hyperbolic law of cosines
\begin{equation}\label{eq:x}
\cosh x_{ab} = \cosh r_a \cosh r_b - \sinh r_a \sinh r_b
\cos\Delta\theta_{ab},
\end{equation}
where $\Delta\theta_{ab}$ is the angle between $\overline{Oa}$ and
$\overline{Ob}$. (The same formula with $r_O=0$ can be used to
compute $x_{Oa}=r_a$ and $x_{Ob}=r_b$.) Upon distributing nodes over
the disc as described, we form scale-free networks in the model by
connecting each pair of nodes $i$ and $j$ located at hyperbolic
distance $x_{ij}$ with the connection probability
\begin{equation}\label{eq:p(x)}
p(x_{ij})=\left(1+e^{\frac{x_{ij}-R}{2T}}\right)^{-1},
\end{equation}
almost identical to the Fermi-Dirac distribution in statistical
mechanics. It depends only on hyperbolic distances $x_{ij}$ (link
energies), hyperbolic disc radius $R$ (chemical potential), and
parameter $T\geq0$ (temperature) controlling network clustering.
This connection probability results in average node degrees
exponentially decreasing with the distance from the origin, which we
also observe in Fig.~\ref{fig:1}. The combination of an
exponentially increasing node density and exponentially decreasing
average degree yields a power-law node degree distribution in the
network. See Appendix~\ref{model} for further details.

\section{The mapping method}

As our goal is to build a realistic Internet map, ready for routing
and other applications, we have to find for each AS its radial and
angular coordinates $(r,\theta)$ maximising the efficiency of greedy
forwarding. This specific task of maximising greedy
forwarding efficiency calls for a mapping method different from
existing techniques on embedding Internet distances and
graphs~\cite{TaCo03,ShaTa04a,ShaTa08}. In view of our previous
findings~\cite{BoKrKc08,BoKr09,KrPaBo09,PaKr10} that greedy
forwarding is exceptionally efficient in Internet-resembling
synthetic networks, and that this efficiency is maximised in the
Einsteinian model, our strategy for the Internet map construction is
to maximise the congruency between the map and the model. In
statistical inference~\cite{Cox05-book}, this goal is equivalent to
maximising the likelihood that the observed data, i.e., the Internet
topology, has been produced by the model. This likelihood is given
by
\begin{equation}
\label{eq:likelihood} {\cal L}=\prod_{i<j}
p(x_{ij})^{a_{ij}}[1-p(x_{ij})]^{1-a_{ij}},
\end{equation}
where the elements $a_{ij}$ of the Internet adjacency matrix are
equal to $1$ whenever there exists a connection between ASs $i$ and
$j$, and to $0$ otherwise. While the adjacency matrix represents the
observed data, the connection probability $p(x_{ij})$ depends via
Eqs.~(\ref{eq:p(x)},\ref{eq:x}) on the AS coordinates $(r,\theta)$,
which we try to infer. Our best estimate for these coordinates are
then those maximising the likelihood in Eq.~(\ref{eq:likelihood}).

\begin{figure*}
\centerline{\includegraphics[width=7in]{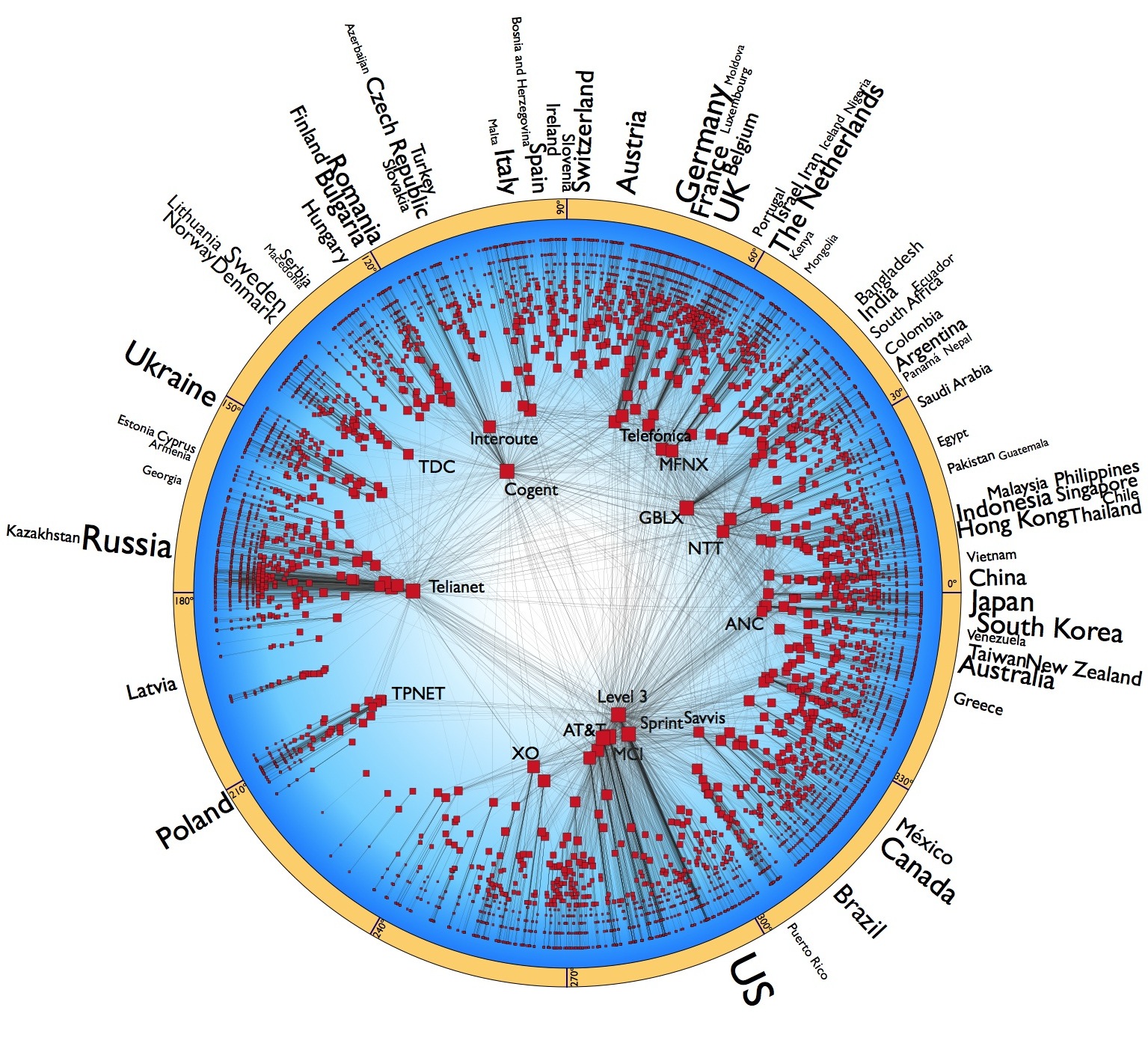}}
\caption{The hyperbolic map of the Internet is similar to a
synthetic Einsteinian network in Fig.~\ref{fig:1}. The size of AS
nodes is proportional to the logarithm of their degrees. For the
sake of clarity, only ASs with degree above $3$, and only the
connections with probability $p(x)>0.5$ given by Eq.~(\ref{eq:p(x)})
are shown. The font size of the country names is proportional to the
logarithm of the number of ASs that the country has. Only the names
of countries with more than $10$ ASs are included. The methods used
to map ASs to their countries are described in Appendix~\ref{appendixG}.
\label{fig:2}}
\end{figure*}

Although there are a plenty of methods to find maximum-likelihood
solutions, e.g., the Metropolis-Hastings
algorithm~\cite{Newman99-book}, they perform poorly and do not scale
well on large datasets with abundant local maxima, which is the case
with the Internet. Therefore, as important as a maximisation method
is a heuristic approach helping the maximisation algorithm to find
the optimal solution in a reasonable amount of time and with
reasonable computational resources. Our method is based on the
following remarkable property of networks in our model; the same
property holds for the Internet~\cite{SeKrBo08}. Let ${\cal G}$ be a
given network with average degree $\bar{k}$ and power-law degree
distribution $P(k) \sim k^{-\gamma}$, and let ${\cal G}(k_T)$ be
${\cal G}$'s subgraph composed of nodes with degree larger than some
threshold $k_T$, along with the connections among these nodes. The
average degree in ${\cal G}(k_T)$ is then given by
$\bar{k}(k_T)=k_T^{3-\gamma}\bar{k}$~\cite{SeKrBo08}. In scale-free
networks with exponent $\gamma$ between $2$ and $3$, this internal
average degree is thus a growing function of $k_T$, which implies
that subgraphs made of high degree nodes almost surely form a single
connected component. Using this property along with the statistical
independence of the graph edges, it becomes possible to infer
coordinates of ASs in ${\cal G}(k_T)$ ignoring the remainder of the
AS graph. This property is practically important because the size of
${\cal G}(k_T)$ decreases very fast as $k_T$ increases, which speeds
up likelihood maximisation algorithms tremendously. In a nutshell,
our method starts with a subgraph ${\cal G}(k_T)$ small enough for
standard maximisation algorithms being able to reliably and quickly
infer the coordinates of ASs in ${\cal G}(k_T)$. Once these are
found, we gradually increase $k_T$ to iteratively add layers of
lower-degree ASs. While doing so, we use the already inferred AS
coordinates as a reference frame to assign initial coordinates to
newly added ASs. This initial coordinate assignment significantly
improves the convergence time of maximisation algorithms. All other
details of our mapping method can be found in Appendix~\ref{appendixC}.

\section{Mapping results}

\begin{figure}
\centerline{\includegraphics[width=3.5in]{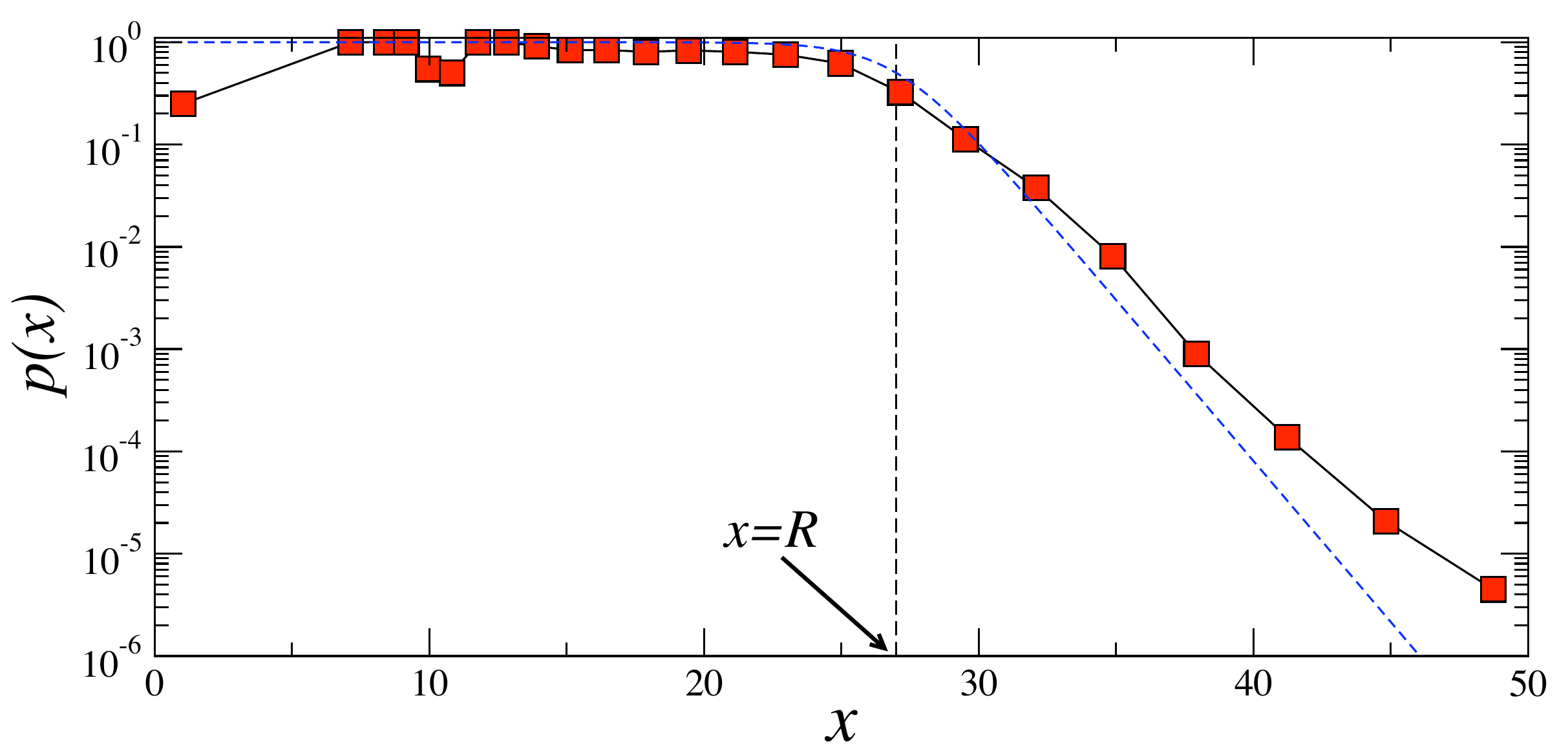}}
\caption{Hyperbolic mapping of the Internet is successful, as the
empirical connection probability between ASs of degree larger than
$2$ in the map closely follows the Einsteinian model prediction. The
whole range of hyperbolic distances $x\in[0,2R]$ is binned, and for
each bin the ratio of the number of connected AS pairs to the total
number of AS pairs falling within this bin is shown. The distances
between AS pairs are computed using Eq.~(\ref{eq:x}). The blue
dashed line is the connection probability given by
Eq.~(\ref{eq:p(x)}) with $R=27$ and $T=0.69$, which are the values
used by the mapping method. \label{fig:3}}
\end{figure}

\begin{figure}
\centerline{\includegraphics[width=3.5in]{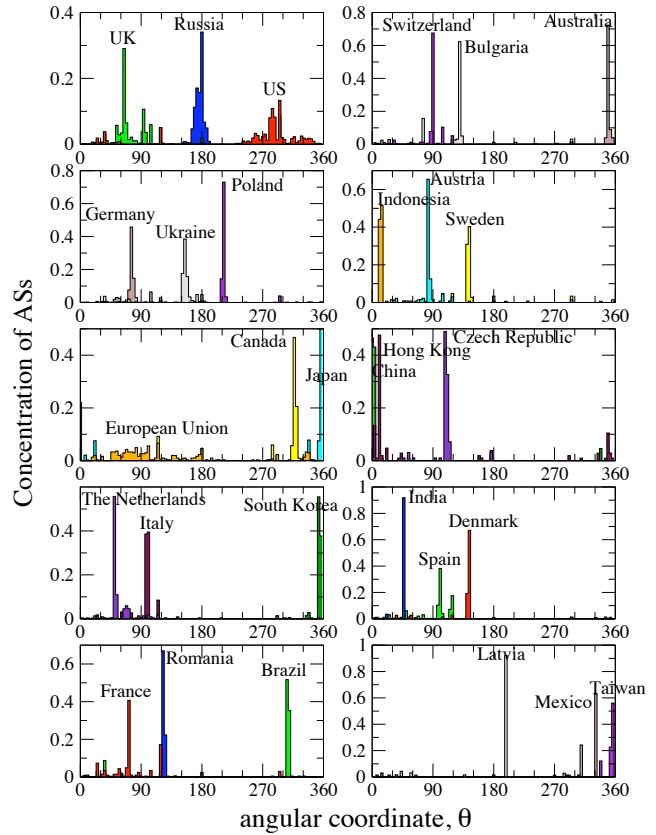}}
\caption{Hyperbolic mapping of the Internet yields meaningful
results, as ASs belonging to the same country are mapped close to
each other. The angular distributions of ASs in the thirty largest
countries in the world are shown. The ``size'' of the country is the
number of ASs it has. The graph shows the percentage of ASs per bin
of size $3.6^o$. For the majority of countries, their ASs are
localised in narrow regions. The exceptions are the US, EU, and UK.
The first two exceptions are due the significant geographic spread
of ASs belonging to the US or EU, the latter actually representing
not one country but a collection of countries. \label{fig:4}}
\end{figure}

We apply our mapping method to the Internet AS topology extracted
from the Archipelago project data~\cite{ClHy09-short} in June 2009
and described in Appendix~\ref{ark}, and
visualise the results in Fig.~\ref{fig:2}. We observe striking
similarity between this visualisation and the synthetic Einsteinian
network in Fig.~\ref{fig:1}. To confirm that the Internet map we
have obtained is indeed congruent with the Einsteinian model, we
juxtapose in Fig.~\ref{fig:3} the empirical connection probability
between ASs in the obtained Internet map against the theoretical one
in Eq.~(\ref{eq:p(x)}). We observe a clear similarity between the
two. Neither the sphere is a perfect model of the Earth, nor the
Einsteinian model is an ideal abstraction of the Internet structure.
Yet, the observed similarity between the empirical and theoretical
connection probabilities in Fig.~\ref{fig:3} suggests that
hyperbolic metric spaces coupled with Fermi-like connection
probabilities are reasonable representations of the real Internet
space.

To investigate further the connections between the obtained map and
Internet reality, we show in Fig.~\ref{fig:2} the average angular
position of all ASs belonging to the same country, while in
Fig.~\ref{fig:4} we draw the angular distributions of those ASs.
Surprisingly, we find that even though our mapping method is
completely geography-agnostic, it discovers meaningful groups or
communities of ASs belonging to the same country. Furthermore, in
Fig.~\ref{fig:2} we find many cases of geographically or politically
close countries placed close to each other in our hyperbolic map.
The explanation of these surprising effects is rooted in the
peculiar nature of our mapping method. If ASs belonging to the same
country, geographic region, or geo-political or
economic group are connected more densely to each other than to the
rest of the world, then this higher connection density translates to
a higher attractive force that tries to place all such ASs close to
each other in our map. Indeed, the term $p(x_{ij})^{a_{ij}}$ in
Eq.~(\ref{eq:likelihood}) corresponds to the attractive force
between connected nodes, while the term $[1-p(x_{ij})]^{1-a_{ij}}$
is the repulsive force between disconnected ones. This peculiar
interplay between attraction within densely connected regions, and
repulsion across sparsely connected zones, effectively maps closely
the ASs belonging to densely connected AS groups. These observations
build our confidence that our mapping method provides meaningful
results reflecting peculiarities of the real Internet structure, and
suggest that the method can be adapted to discover the community
structure~\cite{GiNe02,Newman06,CaVe07} in other complex networks.

\section{Routing results}

The obtained Internet map is ready for greedy forwarding. An AS
holding a packet reads its destination AS coordinates, computes the
hyperbolic distances between this destination and each of its AS
neighbours using Eq.~(\ref{eq:x}), and forwards the packet to the
neighbour closest to the destination. To evaluate the performance of
this process, we perform greedy forwarding from each
source to each destination AS, and compute several performance
metrics.

The first metric is success ratio, which is the percentage of greedy
paths that successfully reach their destinations. Not all paths are
expected to be successful as some might run into local minima. For
example, an AS might forward a packet to its neighbour who sends the
packet back to the same AS, in which case the packet will never
reach the destination. We declare a path unsuccessful, if the packet
is sent to the same AS twice. The average success ratio of simple
greedy forwarding in our Internet map is remarkably high, $97\%$,
and more sophisticated greedy forwarding techniques, such as those
described in~\cite{CvCr09}, can boost it to $100\%$.

Given the
discussed connections between our Internet map and geography, one
may conjecture that greedy forwarding simply mimics geographic
routing following the geographically shortest paths. However, this
conjecture is not true. Geography is reflected in our map only along
the angular coordinate, while the radial coordinate is a function of
the AS degree, making the space hyperbolic, see Appendix~\ref{model}.
The geographic space is not hyperbolic, and if we
use it for greedy forwarding, we obtain a much lower success ratio
of approximately $14\%$. We also tested modified geographic routing
that tries to intelligently use AS degrees, in spirit of our
Einsteinian model. Nevertheless, this modification, although
improving the success ratio to $30\%$, still fails short compared to
the results obtained using our hyperbolic map. The details of these
experiments with geographic routing are in Appendix~\ref{appendixH}.

The second metric is stretch, which tells us by how much longer the
greedy paths are, compared to shortest paths in the Internet
topology. The average stretch is low, $1.1$. The average hop-wise
length of the shortest paths between selected sources and
destinations is $3.49$, so that the average length of greedy paths
is $3.86$. The low value of stretch indicates that
greedy paths are close to optimal, i.e., shortest paths. The
shortest path between nodes $a$ and $b$ in Fig.~\ref{fig:1}, for
example, is also the path found by greedy forwarding. Somewhat
unexpectedly, the greedy stretch is asymptotically optimal, i.e.,
equal to $1$, in scale-free, strongly clustered networks regardless
what underlying space is used for greedy forwarding~\cite{BoKr09}.
Low stretch also implies that greedy forwarding causes approximately
the same traffic load on nodes as shortest-path forwarding. Given
that shortest-path forwarding does not lead to high traffic load in
scale-free networks~\cite{GkaMiSa03}, this finding allays concerns
that hyperbolic forwarding may cause traffic congestion
abnormalities~\cite{JoLo10}. More details on this topic are in
Appendix~\ref{appendixL}.

\begin{figure}
\centerline{\includegraphics[width=3.5in]{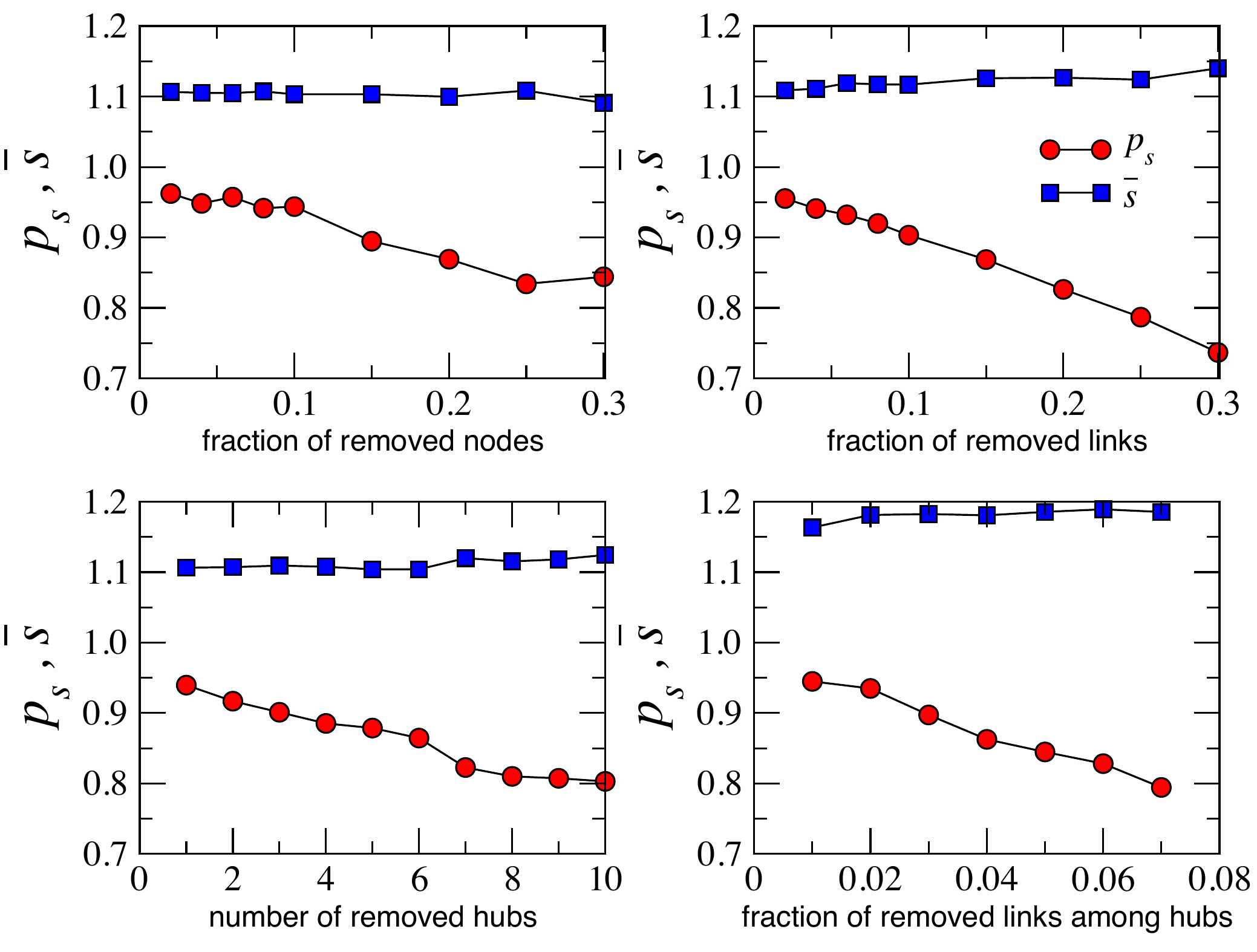}}
\caption{Greedy forwarding performs almost optimally in the mapped
Internet, as indicated by the success ratio, $p_s$, and average
stretch, $\bar{s}$, after removal of a given fraction of AS nodes
(top left) or links (top right).
The bottom plots show these two metrics after removing
a number of the highest-degree nodes (bottom left), and a fraction
of links among highest-degree nodes (bottom right). The links are
first ranked by the product of node degrees that they connect, and
then a fraction of top-ranked links are removed. The giant connected
component is still present after all removals, but it drops to
$85\%$ of the original graph after the removal of $10$ hubs.
\label{fig:6}}
\end{figure}

The two metrics above characterise the performance of greedy
forwarding in the static Internet topology. More important than that
is how greedy forwarding performs in the dynamic topology, where
links and nodes can fail. We randomly select a percentage of links
and nodes, remove them from the mapped Internet, recompute the
success ratio and stretch after the removal, and present the result
in the top plots of Fig.~\ref{fig:6}. Even upon
simultaneous failures of up to $10\%$ of AS links or
nodes---catastrophic events never happened in the Internet
history---we observe only minor degradation of the performance of
greedy forwarding. That is, even catastrophic levels of damage to
the Internet does not significantly affect the performance of greedy
forwarding, even though no AS changes its position on the hyperbolic
map.

A widely popularized feature of complex networks
is their robustness with respect to random failures, and the
lethality of failures of highest-degree
hubs~\cite{AlJeBa01,JeMaBaOl01}. As expected we observe in the
bottom plots of Fig.~\ref{fig:6} that removals of such hubs have a
more detrimental effect on greedy forwarding as well. However,
targeted removal of highest-degree ASs in the Internet is a rather
unrealistic scenario since these large ASs consist of thousands of
routers whose simultaneous failure is a very rare and unlikely
event. The explanation for the surprising efficiency of greedy
forwarding with respect to random failures lies in the unique
combination of the following two properties exhibited by scale-free,
strongly clustered networks: high path diversity~\cite{GkaMiSa03},
and congruency between hyperbolic geodesics and topologically
shortest paths~\cite{PaKr10,KrPa09,KrPa10}. The latter is illustrated by
the similar path patterns of the hyperbolic geodesic and
topologically shortest path between nodes $a$ and $b$ in
Fig.~\ref{fig:1}: they both first go to the high-degree core of the
network, and then exit it in the appropriate direction to the
destination. Due to high path diversity, there are many disjoint
shortest paths between the same source and destination, and thanks
to the congruency, they all stay close to the corresponding
hyperbolic geodesics. Link and node failures affect some shortest
paths, but others remain, and greedy forwarding can still find them
using the same hyperbolic map.

Another form of Internet dynamics is its rapid
growth over years~\cite{DhDo08-short,Carpenter09,PaSaVe04-book,CrKr06-book}.
We show in Appendix~\ref{appendixD} that if the
existing ASs keep their hyperbolic coordinates fixed, while the ASs
joining the Internet anew compute their coordinates using local
information, then the performance of greedy forwarding does not
significantly degrade, even at long time scales. In a nutshell, the
existing AS coordinates are essentially static, as they can stay the
same for years.

Existing Internet topology measurements including
the Archipelago data~\cite{ClHy09-short} are known to be incomplete and
miss some AS links. Therefore a natural question is how this missing
information affects the quality of the constructed map, and the
performance of greedy forwarding in it. Intuitively, since the
performance of greedy forwarding is robust with respect to link
removals, then we might expect it to be robust with respect to
missing links as well. Moreover, if the constructed map is used in
practice, then greedy forwarding will see and use those links that
topology measurements do not see. We might thus also intuitively
expect greedy forwarding to perform better in practice than we
report in this section, simply because those missing links, when
used by greedy forwarding, would provide additional shortcuts
between potentially remote ASs. We confirm this intuition in
Appendix~\ref{appendixE} with experiments emulating the missing
link issue. Therefore the routing results reported here should
actually be considered as lower bounds for greedy routing
performance that can be achieved in practice using the constructed
hyperbolic Internet map.

\section{Conclusion}

We have constructed a hyperbolic map of the Internet, and release
this map with this paper~\cite{BoPa10-SI}. The map can be used for
essentially infinitely scalable Internet routing. The amount of
routing information that ASs must maintain is proportional to the AS
degree, which is theoretically best possible since ASs must always
keep some information about their neighbours. Routing communication
overheads are also minimised, since ASs do not exchange any routing
information upon dynamic changes of the AS topology. The presented
solution thus achieves routing efficiency close to theoretically
optimal, and resolves serious scaling limitations that the Internet
faces today.

The mapping method we have employed is generic, and can be applied
to other complex networks with underlying metric
structures and heterogeneous degree distributions. We showed
in~\cite{SeKrBo08} that a good indicator for the presence of an
underlying metric structure is self-similarity of clustering in the
network, while in~\cite{KrPa09,KrPa10} we showed that as soon as a metric
space is present, and the network has a heterogeneous degree
distribution, the metric distances can be rescaled such that the
underlying geometry is effectively hyperbolic. Roughly,
self-similar clustering is responsible for the metric structure
along the angular coordinate, while degree heterogeneity adds the
radial dimension, and makes the space hyperbolic. Applied to other
networks, our mapping method can provide a different perspective on
the community structure in networks. Instead of trying to split
nodes into discrete community sets~\cite{GiNe02,Newman06,CaVe07}, it
would naturally yield a continuous measure of similarity between
nodes based on hyperbolic distances. More similar nodes would be
located closer to each other, and form zones of higher connectivity
density. It would be then up to an experimenter to define
communities, if needed, as histograms of the node density in the
hyperbolic space. The spectrum of potential applications of this
network-mapping geometrisation agenda is wide. Network mapping can
reveal geometric forces effectively driving information signaling in
the network; examples include the brain~\cite{BuSp09} and cell
signaling networks~\cite{Charlebois09}. One can then potentially
predict what network perturbations drive these networks to failures,
such as brain disorders or cancer. Other applications range from
recommender systems~\cite{Monroe09}, where to have the right measure
of similarity between consumers is a key, to epidemic
spreading~\cite{MeAr09} and information theory of
networks~\cite{BiPi09}.

We have shown that the Internet hyperbolic map is remarkably robust
with respect to even substantial perturbations of the Internet
topology, implying that this map is essentially static. It does not
significantly depend on topology dynamics, and can thus be computed
only once. This property is desirable in view of long running times
intrinsic to likelihood maximisation algorithms. Our method improves
their running times drastically, and the Internet map computations
take approximately a day on a modern computer. However, for
substantially larger networks the running times may still be
prohibitive even for one-time mapping. Therefore, alternative
methods for network mapping, not relying on likelihood maximisation,
are highly desirable, and our work in this direction is underway.

\appendix

\section{The Einsteinian and Newtonian models of complex networks}\label{model}
To synthesise a network with our Einsteinian model, one has first to
specify any desired network size $N$, average degree $\bar{k}$,
average clustering $\bar{C}$, and exponent $\gamma>2$ of the
power-law distribution $P(k)$ of node degrees $k$, $P(k) \sim
k^{-\gamma}$. Equipped with these target properties of the network
topology, we first distribute quasi-uniformly $N$ nodes within a
hyperbolic disc of radius $R=2\log(N/c)$, where $c$ is given by
\begin{equation}
c = \bar{k}\frac{\sin\pi T}{2T}
\left(\frac{\gamma-2}{\gamma-1}\right)^2,
\end{equation}
and $T\in[0,1]$ is a function of $\bar{C}$. In the hyperbolic plane,
the quasi-uniform node density means that the node angular
coordinates $\theta\in[0,2\pi]$ are distributed uniformly, while
their radial coordinates $r\in[0,R]$ are distributed with density
\begin{equation}\label{eq:alpha}
\rho(r) = \alpha e^{\alpha(r-R)},
\end{equation}
where $\alpha=(\gamma-1)/2$. Once all nodes are in place specified
by their assigned coordinates, the hyperbolic distance $x_{ij}$
between each pair of nodes $i$ and $j$ located at $(r_i,\theta_i)$
and $(r_j,\theta_j)$ is computed using Eq. (\ref{eq:x}). Given these
distances, each pair of nodes $i$ and $j$ is then connected by a
link with probability $p(x_{ij})$ given by Eq.~(\ref{eq:p(x)}).
After each node pair is examined and connected with probability
$p(x_{ij})$, the network is formed, and we can compute the average
degree $k(r)$ of nodes located at distance $r$ from the origin. The
result is
\begin{equation}\label{eq:k(r)}
k(r) = \bar{k}\frac{\gamma-2}{\gamma-1} e^{(R-r)/2},
\end{equation}
which combined with Eq.~(\ref{eq:alpha}) yields the target degree
distribution $P(k)$. The Newtonian model is isomorphic to the
Einsteinian one via a simple change of variables reminiscent to
Eq.~(\ref{eq:k(r)}):
\begin{equation}\label{eq:kappa.vs.r}
\kappa = \kappa_0 e^{(R-r)/2},
\end{equation}
where $\kappa$ is the expected degree of a node in the Newtonian
model, and $\kappa_0$ is the minimum expected degree.
See~\cite{KrPa09,KrPa10} for further details.

\section{Mapping methods}\label{sec:AS-embedding}
\label{appendixC}

To find our hyperbolic Internet map, we use the equivalence between
the Einsteinian--$\mathbb{H}^2$~\cite{KrPa09,KrPa10} and the
Newtonian--$\mathbb{S}^1$~\cite{SeKrBo08} models. This equivalence
establishes a relationship in Eq.~(\ref{eq:kappa.vs.r})
between the expected degree $\kappa$ of a
node in the Newtonian--$\mathbb{S}^1$ model, and its radial
coordinate $r$ in the Einsteinian--$\mathbb{H}^2$ model
The angular coordinate $\theta$ is the same in both
models. Thus, for a given node $i$ we aim to find its expected
degree and angular coordinate, $\{\kappa_i,\theta_i\}$, that best
match the Newtonian--$\mathbb{S}^1$ model. We then use the
$\kappa$-to-$r$ mapping to place nodes in the hyperbolic plane
according to the Einsteinian--$\mathbb{H}^2$ model.

Thanks to their equivalence, the Newtonian--$\mathbb{S}^1$ and
Einsteinian--$\mathbb{H}^2$ models generate statistically the same
network topologies. However, the efficiency of greedy forwarding in
the Einsteinian--$\mathbb{H}^2$ model is higher, because hyperbolic
geodesics are exceptionally congruent with the topologically
shortest paths in scale-free networks~\cite{KrPa09,KrPa10,KrPaBo09}. The
reason for this congruency is that the \emph{effective} distance
used as an argument of the connection probability in the
Newtonian--$\mathbb{S}^1$ is actually hyperbolic~\cite{KrPa09,KrPa10}, and
the Einsteinian--$\mathbb{H}^2$ model simply translates this
effective distance to the real hyperbolic one. For these reasons we
prefer the Einsteinian--$\mathbb{H}^2$ model for routing purposes,
although we use the Newtonian--$\mathbb{S}^1$ one to find the
Internet map. We could use directly the Einsteinian--$\mathbb{H}^2$
model for this purpose, but the Newtonian--$\mathbb{S}^1$ model is
technically simpler since the statistical inference in it can be
performed independently for the two variables $\kappa$ and~$\theta$.

We first recall the Newtonian--$\mathbb{S}^1$ model, which generates
networks according to the following steps:
\begin{enumerate}
\item
Distribute $N$ nodes uniformly over the circle $\mathbb{S}^1$ of
radius $N/(2\pi)$, so that the node density on the circle is fixed
to $1$~\footnote{We chose the uniform distribution because we do not
have any {\it a priori} preferred angular coordinate values, and
thus expect the network to be isotropic.}.
\item
Assign to all nodes a hidden variable $\kappa$ representing their
expected degrees. To generate scale-free networks, $\kappa$ is drawn
from the power-law distribution
\begin{eqnarray}
\rho(\kappa)&=&\kappa_0^{\gamma-1}(\gamma-1)\kappa^{-\gamma},\quad\kappa\in[\kappa_{0},\infty),\\
\kappa_0&=&\bar{k}\frac{\gamma-2}{\gamma-1},\label{eq:kappa0_fixed}
\end{eqnarray}
where $\kappa_0$ is the minimum expected degree, and $ \bar{k}$
is the network average degree~\footnote{Note that the model
generates nodes of zero degree that contribute to the total average
degree.}.
\item
Let $\kappa$ and $\kappa'$ be the expected degrees of two nodes
located at distance $d=N\Delta\theta/(2\pi)$ measured over the
circle, where $\Delta\theta$ is the angular distance between the
nodes. Connect each pair of nodes with probability $p(\chi)$, where
the \emph{effective} distance $\chi \equiv d/(\mu \kappa \kappa')$,
and $\mu$ is a constant fixing the average degree.
\end{enumerate}
The connection probability $p(\chi)$ can be any integrable function.
Here we chose the Fermi-Dirac distribution
\begin{equation}
p(\chi)=\frac{1}{1+\chi^\beta},
\end{equation}
where $\beta=1/T$ is a parameter that controls clustering in the
network. With this connection probability, parameter $\mu$ becomes
\begin{equation}
\mu=\frac{\beta}{2\pi \bar{k}} \sin{\left[\frac{\pi}{\beta}\right]}.
\end{equation}
The expected degree of a node with hidden variable $\kappa$ is
$\bar{k}(\kappa)=\kappa$ and, therefore, the degree distribution
scales as $P(k)\sim k^{-\gamma}$ for large $k$.

To go from the Newtonian--$\mathbb{S}^1$ to the
Einstenian--$\mathbb{H}^2$ models, we leave the angular coordinate
$\theta$ unchanged, while the radial coordinate of a node with
expected degree $\kappa$ is given by
\begin{equation}
r=R-2 \ln{\frac{\kappa}{\kappa_0}},
\end{equation}
where the radius of the hyperbolic disk containing all nodes is
\begin{equation}
R=2 \ln{\left[\frac{N}{\pi \mu \kappa_0^2}\right]}.
\end{equation}

\subsection{General theory behind likelihood maximization}

We now fit the real AS graph to the model. Specifically, given the
measured AS graph, we aim to find the set of coordinates
$\{\kappa_i,\theta_i\}$, $i=1,\cdots,N$, that best match the
Newtonian--$\mathbb{S}^1$ model in a statistical sense. To do so, we
use maximum likelihood estimation (MLE) techniques. Let us compute
the posterior probability, or likelihood, that a network given by
its adjacency matrix $a_{ij}$ is generated by the
Newtonian--$\mathbb{S}^1$ model, ${\cal
L}(a_{ij}|\gamma,\beta,\bar{k})$. This probability is
%\begin{widetext}
\begin{equation}
{\cal L}(a_{ij}|\gamma,\beta,\bar{k})=\int \cdots \int {\cal L}(a_{ij},\{\kappa_i,\theta_i\}|\gamma,\beta,\bar{k}) \prod_{i=1}^N d\theta_i d \kappa_i,
\end{equation}
%\end{widetext}
where function ${\cal
L}(a_{ij},\{\kappa_i,\theta_i\}|\gamma,\beta,\bar{k})$ within the
integral is the joint probability that the model generates the
adjacency matrix $a_{ij}$, and the set of hidden variables
$\{\kappa_i,\theta_i\}$. Using Bayes' rule, we find the likelihood
that the hidden variables take particular values
$\{\kappa_i,\theta_i\}$ in the network given by its observed
adjacency matrix $a_{ij}$
%\begin{widetext}
\[
{\cal L}(\{\kappa_i,\theta_i\}| a_{ij},\gamma,\beta,\bar{k})=\frac{{\cal L}(a_{ij},\{\kappa_i,\theta_i\}|\gamma,\beta,\bar{k})}{{\cal L}(a_{ij}|\gamma,\beta,\bar{k})}=
\]
\begin{equation}
=\frac{\mbox{Prob}(\{\kappa_i,\theta_i\}){\cal L}(a_{ij}|\{\kappa_i,\theta_i\},\gamma,\beta,\bar{k})}{{\cal L}(a_{ij}|\gamma,\beta,\bar{k})},
\label{likelihood}
\end{equation}
%\end{widetext}
where
\begin{equation}
\mbox{Prob}(\{\kappa_i,\theta_i\})=\frac{1}{(2\pi)^N}\prod_{i=1}^N \rho(\kappa_i)
\end{equation}
is the prior probability of the hidden variables given by the model,
\begin{equation}
{\cal L}(a_{ij}|\{\kappa_i, \theta_i\}, \gamma,\beta,\bar{k})=\prod_{i<j} p(\chi_{ij})^{a_{ij}} [1-p(\chi_{ij})]^{1-a_{ij}}
\end{equation}
is the likelihood of finding $a_{ij}$ if the hidden variables are
$\{\kappa_i, \theta_i\}$, and
\begin{equation}
\chi_{ij}=\frac{N \bar{k} \Delta \theta_{ij}}{\beta \sin{(\pi/\beta)} \kappa_i \kappa_j},
\end{equation}
\begin{equation}
\Delta\theta_{ij}=\pi-|\pi-|\theta_i-\theta_j||.
\end{equation}
The MLE values of the hidden variables $\{\kappa_i^*, \theta_i^*\}$
are then those that maximize the likelihood in
Eq.~(\ref{likelihood}) or, equivalently, its logarithm,
%\begin{widetext}
\[
\ln{{\cal L}(\{\kappa_i,\theta_i\}| a_{ij},\gamma,\beta,\bar{k})}=
C-\gamma \sum_{i=1}^N \ln{\kappa_i}+
\]
\begin{equation}
+\sum_{i<j} a_{ij} \ln{p(\chi_{ij})}+\sum_{i<j} (1-a_{ij}) \ln{[1-p(\chi_{ij})]},
\label{likelihoodfinal}
\end{equation}
%\end{widetext}
where $C$ is independent of $\kappa_i$ and $\theta_i$.

\subsection{MLE for expected degrees $\kappa$}

The derivative of Eq.~(\ref{likelihoodfinal}) with respect to
expected degree $\kappa_l$ of node $l$ is
%\begin{widetext}
\[
\frac{\partial}{\partial \kappa_l} \ln{{\cal L}(\{\kappa_i,\theta_i\}| a_{ij},\gamma,\beta,\bar{k})}=
\]
\begin{equation}
=-\frac{\gamma}{\kappa_l}-\frac{\beta}{\kappa_l} \left( \sum_{j \ne l} p(\chi_{lj})-\sum_j a_{lj}\right).
\end{equation}
%\end{widetext}
The first term within the parenthesis is the expected degree of node
$l$, while the second term is its actual degree $k_l$. Therefore,
the value $\kappa_l^*$ that maximizes the likelihood is given by
\begin{equation}
\bar{k}(\kappa_l^*)=\kappa_l^*=k_l-\frac{\gamma}{\beta}.
\end{equation}
Since $\kappa_l^*$ can be smaller than $\kappa_0$ in the last
equation, we set
\begin{equation}\label{kappa*}
\kappa_l^*=\max{\left(\frac{\gamma-2}{\gamma-1}\bar{k},k_l-\frac{\gamma}{\beta}\right)}.
\end{equation}
We discuss a correction of this equation accounting for finite size
effects is Section~\ref{finte-size}.

\subsection{MLE for angular coordinates $\theta$}

Having found the MLE values for expected degrees $\kappa$, we now
have to maximize Eq.~(\ref{likelihood}) with respect to angular
coordinates $\theta$. This task is equivalent to maximizing the
partial log-likelihood
%\begin{widetext}
\[
\ln{{\cal L}(a_{ij}|\{\kappa_i^*, \theta_i\}, \gamma,\beta,\bar{k})}=
\]
\begin{equation}
=\sum_{i<j} a_{ij} \ln{p(\chi_{ij})}+\sum_{i<j} (1-a_{ij}) \ln{[1-p(\chi_{ij})]}.
\label{loglike}
\end{equation}
%\end{widetext}
The first term in this equation involves only pairs of connected
nodes, whereas the second term accounts for pairs of disconnected
ones. Since the connection probability $p(\chi)$ is a monotonously
decreasing function of the effective distance $\chi$, the first term
in Eq.~(\ref{loglike}) is large if pairs of connected nodes are
placed close to each other. In contrast, the second term is large if
pairs of disconnected nodes are far apart. Therefore the optimal MLE
solution will balance both effects, and place connected nodes as
close as possible while keeping disconnected ones as far as
possible.

Unfortunately, the maximization of Eq.~(\ref{loglike}) with respect
to the angular coordinates cannot be performed analytically. We thus
have to rely on approximations. At their core lie MLE algorithms, or
kernels, which we discuss first. We present two such kernels,
standard Metropolis-Hastings~(SMH)~\cite{Newman99-book}, and our
``localized'' version of it~(LMH).

\subsubsection{MLE kernels}

In the {\bf standard Metropolis-Hastings~(SMH)} algorithm, a node is
chosen at random, and given a new angular position chosen uniformly
in the interval $[0,2\pi]$. The change is accepted whenever the
likelihood in Eq.~(\ref{loglike}) computed after the change, ${\cal
L}_{new}$~\footnote{From now on we denote the log-likelihood in
Eq.~(\ref{loglike}) by $\ln{\cal L}$.}, is larger than the likelihood
computed with the old coordinate, ${\cal L}_{old}$. Otherwise, the
change is accepted with probability ${\cal L}_{new}/{\cal L}_{old}$.
The SMH algorithm samples the angular phase space, and produces
angular configurations with a probability proportional to the
likelihood. The SMH's computational complexity depends on a
particular system to which SMH is applied. We find that in our case
the number of node moves sufficient for SMH to converge is ${\cal
O}(N^2)$, meaning that the total running time complexity is ${\cal
O}(N^3)$, since each move attempt involves the ${\cal O}(N)$
computation of the likelihood change.

\begin{figure}[t]
\begin{center}
\includegraphics[width=3.5in]{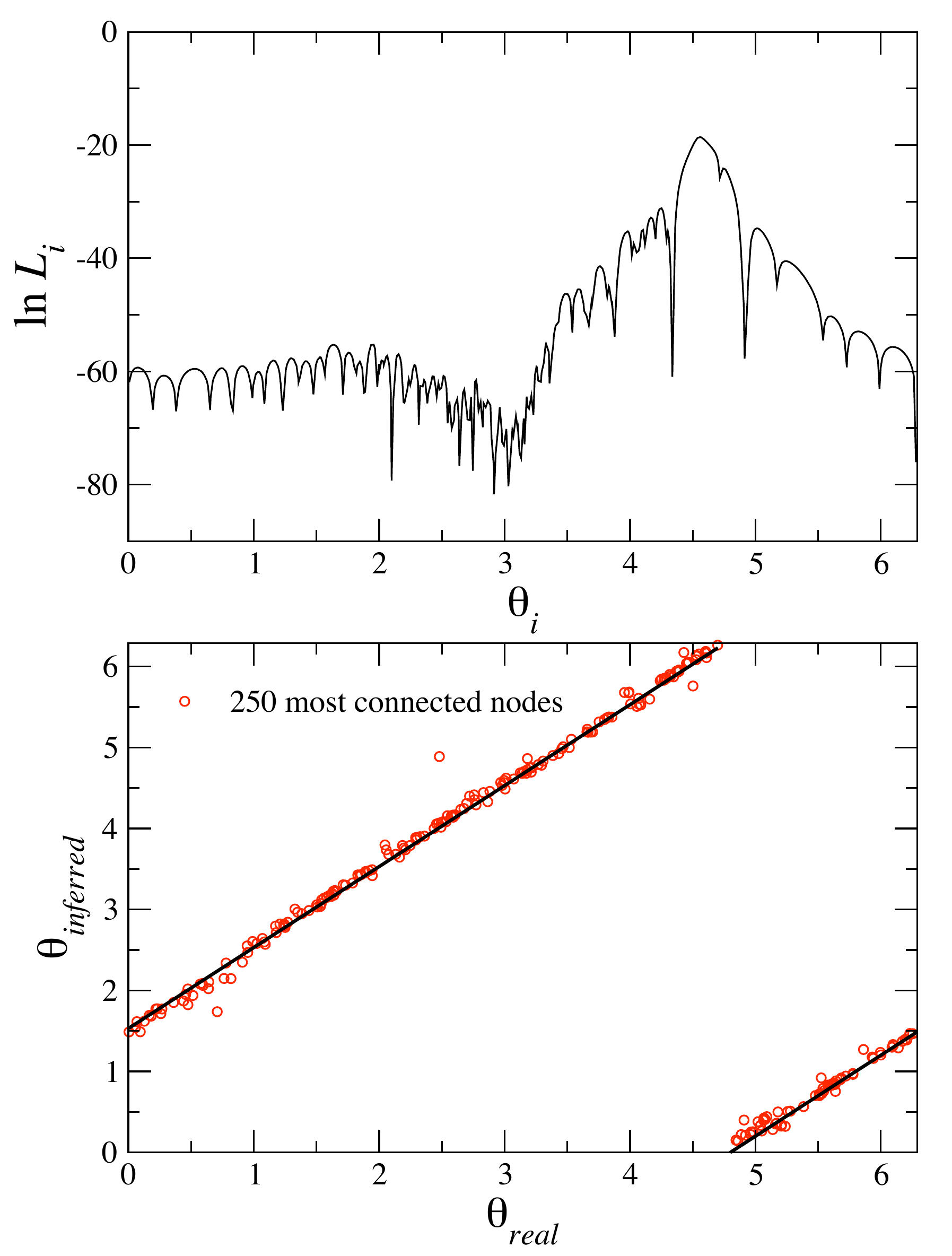}
\caption{Top: example of the local log-likelihood of node $i$
(Eq.~(\ref{loglike-local})), having the coordinates of all other
nodes fixed at an intermediate step of the mapping process. Bottom:
inferred angular coordinates vs.\ real ones for the 250 most
connected nodes in a synthetic $\mathbb{S}^1$ network with the same
properties as the real Internet: $\gamma=2.1$, $\beta=2$, $N \approx
24000$, $\bar{k} \approx 5$. The LMH kernel and
Alg.~\ref{heuristic1} are used for coordinate inference.}
\label{modelembedding}
\end{center}
\end{figure}

Our {\bf localized Metropolis-Hastings~(LMH)} algorithm is not MH
{\it per se}. In fact it bears stronger resemblances to extremal
optimization and genetic search algorithms than to MH. We first
define the local contribution $\ln{{\cal L}_i}$ of node $i$ to the
total log-likelihood $\ln{{\cal L}}$ in Eq.~(\ref{loglike}):
\begin{equation}\label{loglike-local}
\ln{{\cal L}_i}=\sum_{j \ne i} a_{ij} \ln{p(\chi_{ij})}+\sum_{j \ne i} (1-a_{ij}) \ln{[1-p(\chi_{ij})]},
\end{equation}
so that $\ln{{\cal L}}=1/2 \sum_{i} \ln{{\cal L}_i}$. We can
interpret function $\ln{{\cal L}_i}$ as the fitness of node $i$,
which we can then use to maximize the total likelihood.
Specifically, in LMH nodes are visited in rounds, and during each
round all nodes are visited one by one. At each node visit, the node
is moved to the angular position that maximizes its fitness
$\ln{{\cal L}_i}$, having fixed the positions of all other nodes at
that particular node visit. An example of the log-likelihood
landscape that a node sees during its move is shown in the top plot
of Fig.~\ref{modelembedding}. The total number of rounds of all-node
visits needed for LMH to converge is of the order of the network
average degree. Indeed, even though after each node move, the
fitness of other nodes changes, the node fitness is mostly affected
by changes of coordinates of the node neighbors, whose average
number thus roughly determines the number of rounds.
The maximization of the fitness of a node takes ${\cal O}(N^2)$
time, having fitness $\ln{{\cal L}_i}$ sampled at intervals with
$\Delta\theta=1/N$. Therefore for sparse graphs, the overall
computational complexity of LMH is ${\cal O}(N^3)$.

Applied to the real Internet and synthetic Internet-like networks
below, both SMH and LMH yield similar good results. However, we
prefer LMH since by its localized nature, it can be implemented in a
distributed manner, an important property for deployment in the real
Internet. Even more importantly, with LMH, new-coming ASs can
compute their coordinates in a distributed manner {\em without
knowing the global Internet topology}. Indeed, to compute its
coordinates using Eq.~(\ref{loglike-local}), a new-coming AS $i$ has
to know the status of connections only to its neighbors; the status
of connections between any two ASs other than $i$ does not
contribute to $\ln{{\cal L}_i}$ in Eq.~(\ref{loglike-local}). All
results shown in this paper are for LMH.

\subsubsection{First MLE wrapper}

If we na\"{\i}vly applied any MLE kernel to the Internet, we would
have to wait forever for good results. We see in the top plot of
Fig.~\ref{modelembedding} that the characteristic likelihood profile
has abundant local maxima. Therefore an MLE kernel is not guaranteed
to converge to the global maximum in a reasonable amount of time. It
is thus imperative to find a heuristic procedure, i.e., an MLE
wrapper, helping an MLE kernel to find its way towards the global
maximum without being trapped in local maxima. This procedure is
equivalent to using all available information to make an educated
guess of the initial node coordinates.

Our MLE wrapping strategy is based on statistical independence of
edges in our graphs. Thanks to this independence, the coordinates of
a set of nodes can be inferred based only on the partial information
contained in a subgraph formed by the nodes in the set, ignoring the
rest of the network. Consider a small subgraph of the whole network,
for our purposes made of high degree nodes, and remove all nodes and
connections not belonging to this subgraph. Since edges in this
subgraph are statistically independent of other edges, we can
maximize the likelihood corresponding to the subgraph, and infer the
coordinates of the nodes in it based only on this partial
information. If the subgraph is small and dense enough, finding the
optimal MLE solution is easy. Once this solution is found, we can
add more nodes to the network, and use the previously inferred
coordinates as the initial configuration for the new MLE problem.
However, this method works only if the subgraph forms a single
connected component. This property
holds for synthetic networks in our model, and for the real
Internet~\cite{SeKrBo08}.

Formally, let $k_1,k_2,\cdots,k_m$, with $k_1>k_2>\cdots>k_m$, be a
set of predefined degrees, and let ${\cal G}(k_l)$, $l=1,\cdots,m$,
be the subgraphs formed by all nodes of degrees larger or equal to
$k_l$, plus all connections among them. We thus have ${\cal
G}(k_1)\subset {\cal G}(k_2) \subset \cdots \subset {\cal G}(k_m)$,
forming a hierarchy of nested subgraphs. The main idea behind our
MLE wrapper is to run the MLE kernel, either SMH or LMH, in
iterations, starting with the smallest subgraph, and feeding the
coordinates inferred at each iteration to the MLE kernel at the next
iteration.

This idea must be implemented with care. First, subgraph ${\cal
G}(k_1)$ is different from other subgraphs. Indeed, in scale-free
networks, all nodes of degrees larger than $\sim N^{1/2}$ are
connected almost surely. Therefore all such nodes would appear
identical to the MLE kernel, which would thus place them all at the
same location, something that we have to avoid. To solve this
problem, we remove all connections among nodes of degree larger or
equal to $k_1 \sim N^{1/2}$ and start the wrapper algorithm with the
${\cal G}(k_2)$ iteration. Second, iterating from ${\cal G}(k_l)$ to
${\cal G}(k_{l+1})$, we still need to specify the initial
coordinates of the nodes that belong to ${\cal G}(k_{l+1})$ but not
to ${\cal G}(k_{l})~$\footnote{Adding nodes in ${\cal G}(k_{l+1})$
but not in ${\cal G}(k_{l})$, we check if they have at least two
connections to nodes in ${\cal G}(k_{l})$. Otherwise we postpone
introducing such nodes to the first iteration when they start
satisfying this condition.}. While the assignment of random
coordinates to new nodes is possible, it is much more efficient to
try to maximize the likelihood in ${\cal G}(k_{l+1})$ from the very
beginning. In other words, we assign to each new node $i\in{\cal
G}(k_{l+1})\setminus{\cal G}(k_{l})$ the coordinate maximizing
%\begin{widetext}
\begin{equation}\label{likelihood_local}
\ln{{\cal L}_i\left[  {\cal G}(k_{l}) \right]}=
\end{equation}
\[
=\sum_{j \in  {\cal G}(k_l)} a_{ij} \ln{p(\chi_{ij})}+\sum_{j \in  {\cal G}(k_l)} (1-a_{ij}) \ln{[1-p(\chi_{ij})]}.
\]
%\end{widetext}
We note that node $i$ uses information contained only in ${\cal
G}(k_{l})$ to get its initial coordinate. After all new nodes
corresponding to a given iteration are introduced and assigned
initial coordinates, we apply the MLE kernel to the resulting
system. This heuristic MLE wrapping procedure is summarized in
Alg.~\ref{heuristic1}.

\begin{algorithm}[t]
\algsetup{indent=2em}
\caption{First MLE wrapper}
\begin{algorithmic}
\STATE activate nodes in ${\cal G}(k_1)$
\STATE assign random angular coordinates to nodes in ${\cal G}(k_1)$
\STATE remove links among nodes in ${\cal G}(k_1)$
\FOR{$l=2$ to (\# of layers)}
 \FOR{$j=1$ to (\# of nodes in ${\cal G}(k_l)$ not active)}
   \STATE $i \leftarrow$ label of new node in ${\cal G}(k_l)$ not active
    \IF{\# of connections of $i$ with nodes in ${\cal G}(k_{l-1})\ge 2$}
    \STATE activate new node $i$
    \STATE assign to node $i$ coordinate $\theta_i$ maximizing $\ln{{\cal L}_i\left[  {\cal G}(k_{l-1}) \right]}$
     \ENDIF
     \ENDFOR
   \STATE run the MLE kernel on the set of active nodes
\ENDFOR
\end{algorithmic}
\label{heuristic1}
\end{algorithm}

\begin{figure}[t]
\begin{center}
\includegraphics[width=3.5in]{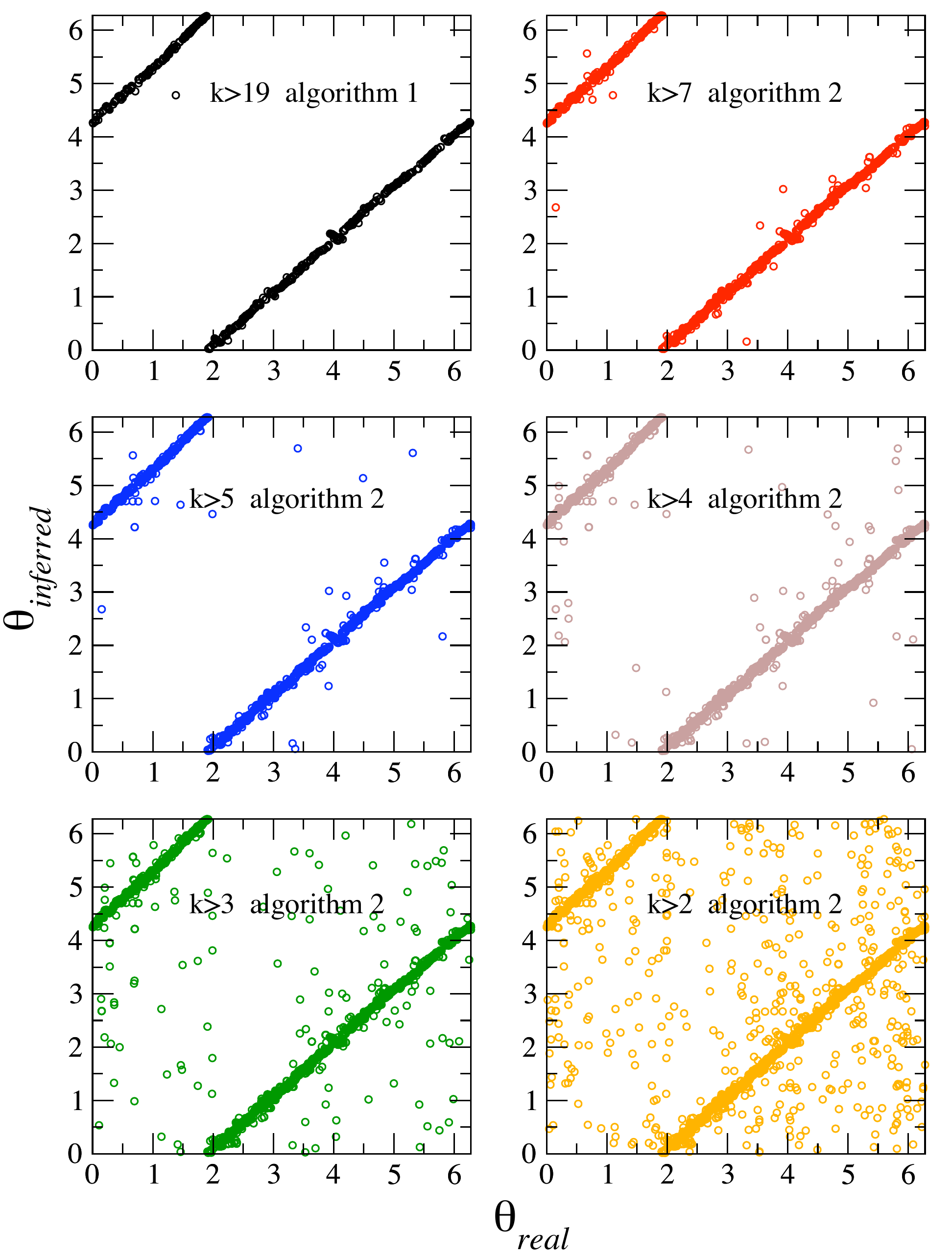}
\caption{Top left: the same as in the bottom plot of
Fig.~\ref{modelembedding} for nodes of degrees $k\ge20$. All other
plots show the results of adding layers of lower degree nodes using
the second MLE wrapper with $k_{l_{critic}}=20$. According to
Alg.~\ref{heuristic2}, nodes in newly added layers infer their
coordinates using the coordinates of nodes in existing layers,
without running the MLE kernel, and the coordinates of existing
nodes do not change as new nodes are added.} \label{modelembedding2}
\end{center}
\end{figure}

In the bottom plot in Fig.~\ref{modelembedding} we show the test
results for this procedure wrapping the LMH kernel, applied to a
synthetic Newtonian--$\mathbb{S}^1$ network generated with the
parameters similar to the real AS graph. We observe that the
inferred coordinates are very close to the real ones, except for a
global phase shift, which can take any value in $[0,2\pi]$ due to
the rotational symmetry of the model.

\subsubsection{Second MLE wrapper}

As mentioned above, it is not necessary to consider the full graph
to infer the coordinates of the most connected nodes. We now use
this observation to speed up the mapping process significantly.
Specifically, we run our first MLE wrapper up to a subgraph of a
certain size, and then add the rest of the nodes assigning to them
their coordinates maximizing Eq.~(\ref{likelihood_local})
\emph{without} subsequent running the MLE kernel, see
Alg.~\ref{heuristic2}.

This modification speeds the overall mapping process because once
the coordinates of the coordinates of a relative small number of
high degree nodes are inferred, the rest of the process takes ${\cal
O}(N^2)$ steps to complete. This improvement reduces the total
running time of the Internet mapping to few hours on a standard
computer. Another practically important feature of this second MLE
wrapper is that new-coming ASs compute their coordinates {\em
without existing ASs changing their coordinates.} In other words,
once the AS coordinates are inferred, they stay static as the
Internet grows.

\begin{algorithm}[t]
\algsetup{indent=2em}
\caption{Second MLE wrapper}
\begin{algorithmic}
\STATE activate nodes in ${\cal G}(k_1)$
\STATE assign random angular coordinates to nodes in ${\cal G}(k_1)$
\STATE remove links among nodes in ${\cal G}(k_1)$
\STATE $l_{critic} \leftarrow$ maximum layer with full MLE calculations
\FOR{$l=2$ to (\# of layers)}
 \FOR{$j=1$ to (\# of nodes in ${\cal G}(k_l)$ not active)}
   \STATE $i \leftarrow$ label of new node in ${\cal G}(k_l)$ not active
    \IF{\# of connections of $i$ with nodes in ${\cal G}(k_{l-1})\ge 2$}
    \STATE activate new node $i$
    \STATE assign to node $i$ coordinate $\theta_i$ maximizing $\ln{{\cal L}_i\left[  {\cal G}(k_{l-1}) \right]}$
     \ENDIF
     \ENDFOR
     \IF{$l \le l_{critic}$}
     \STATE run the MLE kernel on the set of active nodes
     \ENDIF
\ENDFOR
\end{algorithmic}
\label{heuristic2}
\end{algorithm}

\begin{figure}[t]
\begin{center}
\includegraphics[width=3.5in]{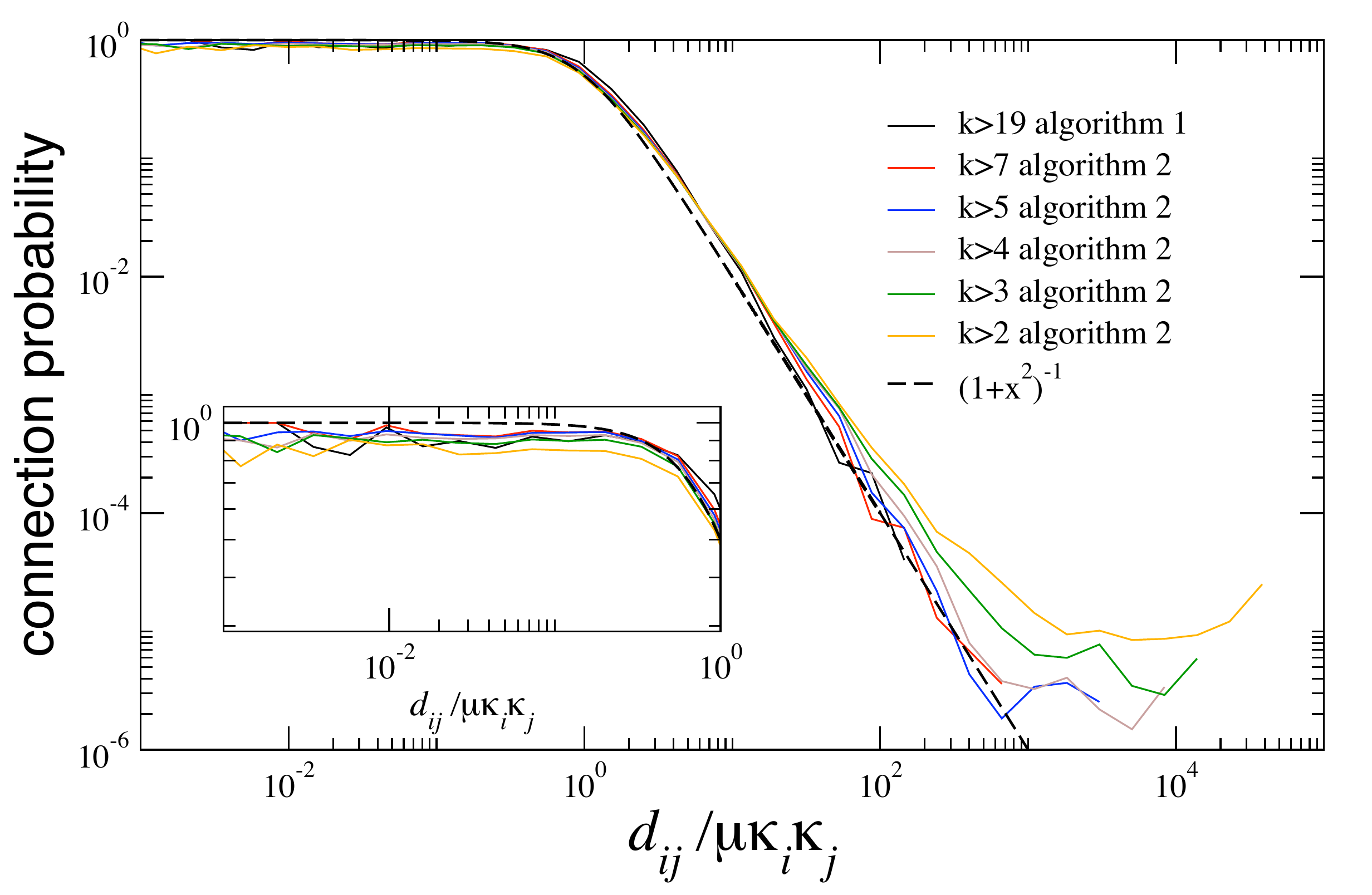}
\caption{Empirical connection probability based on the inferred node
coordinates, compared to the connection probability used to generate
the synthetic network. The inset shows the details in the small
distance region.} \label{modelembedding3}
\end{center}
\end{figure}

We apply this procedure up to nodes of degree $3$. Nodes of degree
$2$ and $1$ must be analyzed separately since all nodes of degree
$1$, and $40\%$ of nodes of degree $2$ do not form any triangles. As
a consequence, the MLE kernel cannot reliably infer their metric
attributes, i.e., their coordinates. Therefore we assign to these
nodes the angular coordinate of their (highest-degree) neighbors,
which makes sense, especially for nodes of degree $1$, since the
only path to such nodes is via their neighbors. Forwarding to such
nodes is thus equivalent to forwarding to their neighbors.

The test results of this second MLE wrapper are shown in
Fig.~\ref{modelembedding2}. The top left plot shows the inferred
vs.\ real coordinates in the same synthetic network for nodes with
degrees $k \ge 20$ using the first MLE wrapper. The other plots show
the corresponding coordinates for nodes with degrees larger than or
equal to $8,6,5,4,3$ using the second MLE wrapper with
$k_{l_{critic}}=20$. That is, the MLE kernel is not run for these
nodes. We observe that the inference quality does deteriorate for
smaller degrees, but it is remarkable that even in the worst case a
majority of coordinates are correctly inferred.

As an additional test, we show in Fig.~\ref{modelembedding3} the
empirical connection probability among nodes in each subgraph using
the coordinates inferred by the second MLE wrapper, compared to the
connection probability $p(x)=(1+x^2)^{-1}$ used to generate the
network. We observe a good agreement for high degree subgraphs,
which slightly deteriorates for low degree nodes located at large
effective distances $\chi$.

To map the AS graph, we used the LMH kernel wrapped with the second
MLE wrapper with $k_{l_{critic}}=20$ and the subgraph hierarchy
defined by $k_1=300$, $k_2=200$, $k_3=160$, $k_4=130$, $k_5=110$,
$k_6=100$, $k_7=90$, $k_8=80$, $k_9=70$, $k_{10}=60$, $k_{11}=50$,
$k_{12}=40$, $k_{13}=30$, $k_{14}=20$, $k_{15}=10$, $k_{16}=9$,
$k_{17}=8$, $k_{18}=7$, $k_{19}=6$, $k_{20}=5$, $k_{21}=4$,
$k_{22}=3$.

\subsection{Parameter estimation and finite size effect}

Our model has three parameters: the exponent $\gamma$ of the degree
distribution, the average degree $\bar{k}$, and the exponent $\beta$
of the connection probability.

\subsubsection{Estimating $\gamma$}

We estimate the exponent $\gamma$ via the direct inspection of the
degree distribution, yielding $\gamma=2.1$.

\subsubsection{Estimating $N$ and $\bar{k}$}\label{finte-size}

The estimation of $N$ and $\bar{k}$ is more involved for two
reasons. First, the Newtonian--$\mathbb{S}^1$ model generates nodes
of zero degree which are included in the computation of the average
degree, $\bar{k}=\sum_{k=0} kP(k)$. However, in the real Internet
graph all nodes have non-zero degrees. Therefore we first have to
estimate the number of nodes $N$ in the model, based on the number
of nodes $N_{obs}$ we observe in the real graph. The relationship
between the two numbers is
\begin{equation}\label{Nfinite}
N=\frac{N_{obs}}{1-P(0)},
\end{equation}
where $P(0)$ is the probability that a node has zero degree in the
model.

The second complication is due to finite size effects. These effects
are particularly important when the exponent $\gamma$ is close to
$2$, which is the case with the Internet. Suppose we generate a
finite size network of $N$ nodes with our Newtonian--$\mathbb{S}^1$
model with parameters $\gamma$, $\bar{k}$, and $\beta$. Since the
network is finite, there is a cut-off value for the expected degree
of a node, $\kappa_c$, which depends on the size of the network. The
first moment of the distribution of expected degrees
$\rho(\kappa)=\kappa_0^{\gamma-1}(\gamma-1)\kappa^{-\gamma}$ with
this cut-off is
\begin{equation}\label{eq:alpha(k,k)}
\langle \kappa (N) \rangle =\bar{k}\left(1-\left[ \frac{\kappa_0}{\kappa_c}\right]^{\gamma-2} \right) \equiv \bar{k} \alpha(\kappa_0,\kappa_c).
\end{equation}
In the thermodynamic limit $\kappa_c\rightarrow \infty$ and
$\alpha(\kappa_0,\kappa_c) \rightarrow 1$. However, if $\gamma$ is
close to $2$, the approach to these limits is slow, and we have to
take care of finite size corrections.

Accounting for these corrections, the expected degree of a node with
hidden variable $\kappa$ becomes
\begin{equation}
\bar{k}_N(\kappa)=\alpha(\kappa_0,\kappa_c) \kappa,
\end{equation}
with $\bar{k}_{\infty}(\kappa)=\kappa$. This equation implies that
the MLE of the hidden variable $\kappa$ of a node of degree $k$
changes from Eq.~(\ref{kappa*}) to
\begin{equation}
\kappa^*=\max{\left(\frac{\gamma-2}{\gamma-1}\bar{k},\frac{1}{\alpha(\kappa_0,\kappa_c)}\left[k-\frac{\gamma}{\beta}\right]\right)},
\end{equation}
while the average degree including zero degree nodes in a finite size
network becomes
\begin{equation}
\bar{k}_N=[\alpha(\kappa_0,\kappa_c)]^2 \bar{k}.
\end{equation}
If the average degree observed in the real Internet graph is
$\bar{k}_{obs}$, our estimate of the parameter $\bar{k}$ is then
\begin{equation}
\bar{k}=\frac{1-P(0)}{[\alpha(\kappa_0,\kappa_c)]^2} \bar{k}_{obs}.
\label{bark}
\end{equation}
Therefore, in order to estimate the values of $N$ and $\bar{k}$ for
a finite network, we first have to estimate the values of $P(0)$ and
$\alpha(\kappa_0,\kappa_c)$. One can check~\cite{SeKrBo08} that
\begin{equation}
P(0)=(\gamma-1) [\alpha(\kappa_0,\kappa_c) \kappa_0]^{\gamma-1} \Gamma(1-\gamma,\alpha(\kappa_0,\kappa_c) \kappa_0),
\label{pzero}
\end{equation}
where $\Gamma(x,y)$ is the incomplete Gamma function. We can also
relate the maximum degree $k^{\max}_{obs}$ observed in the real
Internet to the expected degree cut-off $\kappa_c$ via
\begin{equation}\label{kmax}
k^{\max}_{obs}=\alpha(\kappa_0,\kappa_c) \kappa_c.
\end{equation}

We thus have six unknown values---namely, $N$, $P(0)$, $\kappa_0$,
$\kappa_c$, $\alpha(\kappa_0,\kappa_c)$, and $\bar{k}$---and the
system of six
equations~(\ref{eq:kappa0_fixed},\ref{Nfinite},\ref{eq:alpha(k,k)},\ref{bark},\ref{pzero},\ref{kmax})
involving them. Substituting into these equations the given values
of $N_{obs}=23752$, $\bar{k}_{obs}=4.92$, $k^{\max}_{obs}=2778$, and
$\gamma=2.1$ observed in the Internet, we compute numerically
$\kappa_0=0.9$, $\kappa_c=4790$, $\alpha(\kappa_0,\kappa_c)=0.58$,
and $P(0)=0.33$, yielding $N=35685$ and $\bar{k}=9.86$.

\subsubsection{Estimating $\beta$}

\begin{figure}[t]
\begin{center}
\includegraphics[width=3.5in]{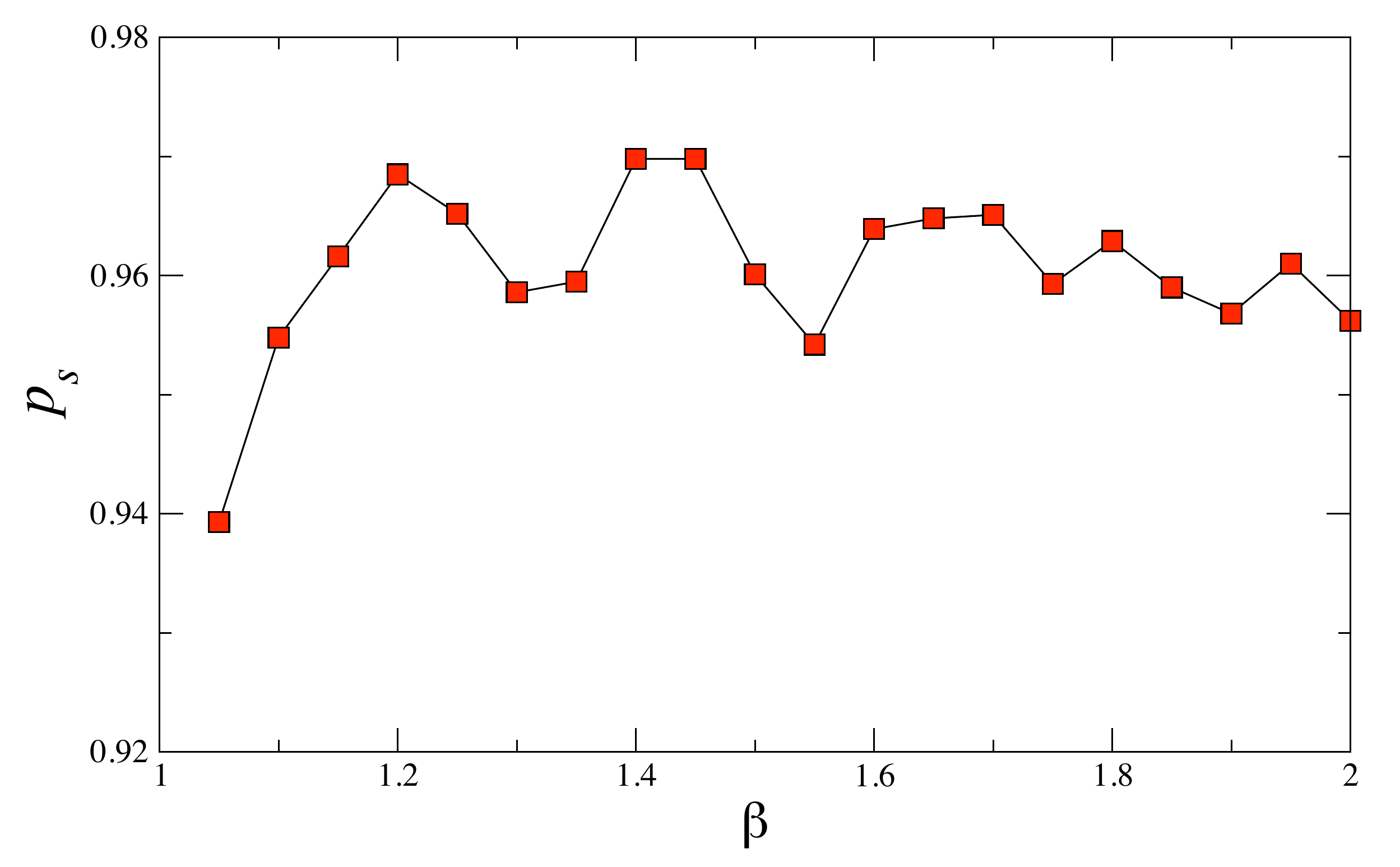}
\caption{Success ratio of greedy forwarding as a function of $\beta$
for the Internet graph mapping.} \label{modelembedding6}
\end{center}
\end{figure}

To estimate $\beta$, we first compare clustering in synthetic
networks with different $\beta$'s to the clustering observed in the
Internet, keeping all other parameters fixed. This procedure narrows
down the possible values of $\beta$ to $\beta \in [1,2]$. We then
generate Internet maps for different values of $\beta$ within this
range, and perform hyperbolic greedy forwarding in them.
Fig.~\ref{modelembedding6} shows the success ratio of greedy
forwarding as a function of $\beta$ in this region. We observe that
the success ratio increases as $\beta$ decreases, and then sharply
drops at $\beta \sim 1$. The value of $\beta$ maximizing the success
ratio is $\beta=1.45$, and we used this value in our final Internet
map.

\section{The Archipelago Internet topology}\label{ark}
We use the AS Internet topology of June 2009 extracted from the data
collected by the Archipelago active measurement infrastructure (ARK)
developed by CAIDA~\cite{ClHy09-short}.
The AS topology contains $23752$ ASs and $58416$ AS links, yielding
the average AS degree $\bar{k}=4.92$. The maximum AS degree is
$k^{\max}=2778$. The average clustering
measured over ASs of degree larger than $1$ is $\bar{C}=0.61$,
yielding temperature $T=0.69$, and hyperbolic disc radius $R=27$.
The exponent of the power-law AS degree distribution is
$\gamma=2.1$. This Internet topology,
along with the hyperbolic Internet map,
are released with this paper~\cite{BoPa10-SI}.

\section{Mapping AS's to countries}
\label{appendixG}
The AS-to-country mapping is taken from the CAIDA AS ranking
project~\cite{DiKrFo06}. It uses two methods for this task. The
first method is {\em IP-based}. It splits the IP address space
advertised by an AS into small blocks, and then maps each block to a
country using~\cite{netacuity}.
If not all IP blocks of an AS map to the same country, then the
other, {\em WHOIS-based} method is used, which reports the country
where the AS headquarters are located according to the WHOIS
database~\cite{whois}. Since large ASs have points of
presence in many countries, they tend to map to multiple countries
using the IP-based method. Therefore, if we did not apply the
WHOIS-based method to them, they would no longer map to a single
country. If we ignored such ASs, the angular distributions of the
remaining ASs belonging to a given country would be even more
localised, including the US, EU, and UK ASs. In our hyperbolic map
data, we release the AS-to-country associations using both methods,
IP+WHOIS-based and IP-based. The latter has no country information
for many ASs with conflicting country mappings.

\section{Geographic routing}
\label{appendixH}
To perform standard geographic routing we first map each AS to a
collection of geographic locations (characterised by their latitudes
and longitudes) using the IP-based method, and then find the centre
of mass for each collection. We thus obtain unique geographic
coordinates for each AS. We then perform standard greedy forwarding
over the AS topology, computing geographic distances between ASs
using the spherical law of cosines. For hyperbolised geographic
routing, we keep the AS angular coordinates equal to their
geographic coordinates, but also, based on the AS degree, we assign
to each AS a radial coordinate, according to the relationship
between node degrees and radial positions in the three-dimensional
Einsteinian model, and then perform greedy forwarding in this
three-dimensional hyperbolic space.

\section{Traffic and congestion considerations}
\label{appendixL}

\begin{figure}[t]
\begin{center}
\includegraphics[width=3.5in]{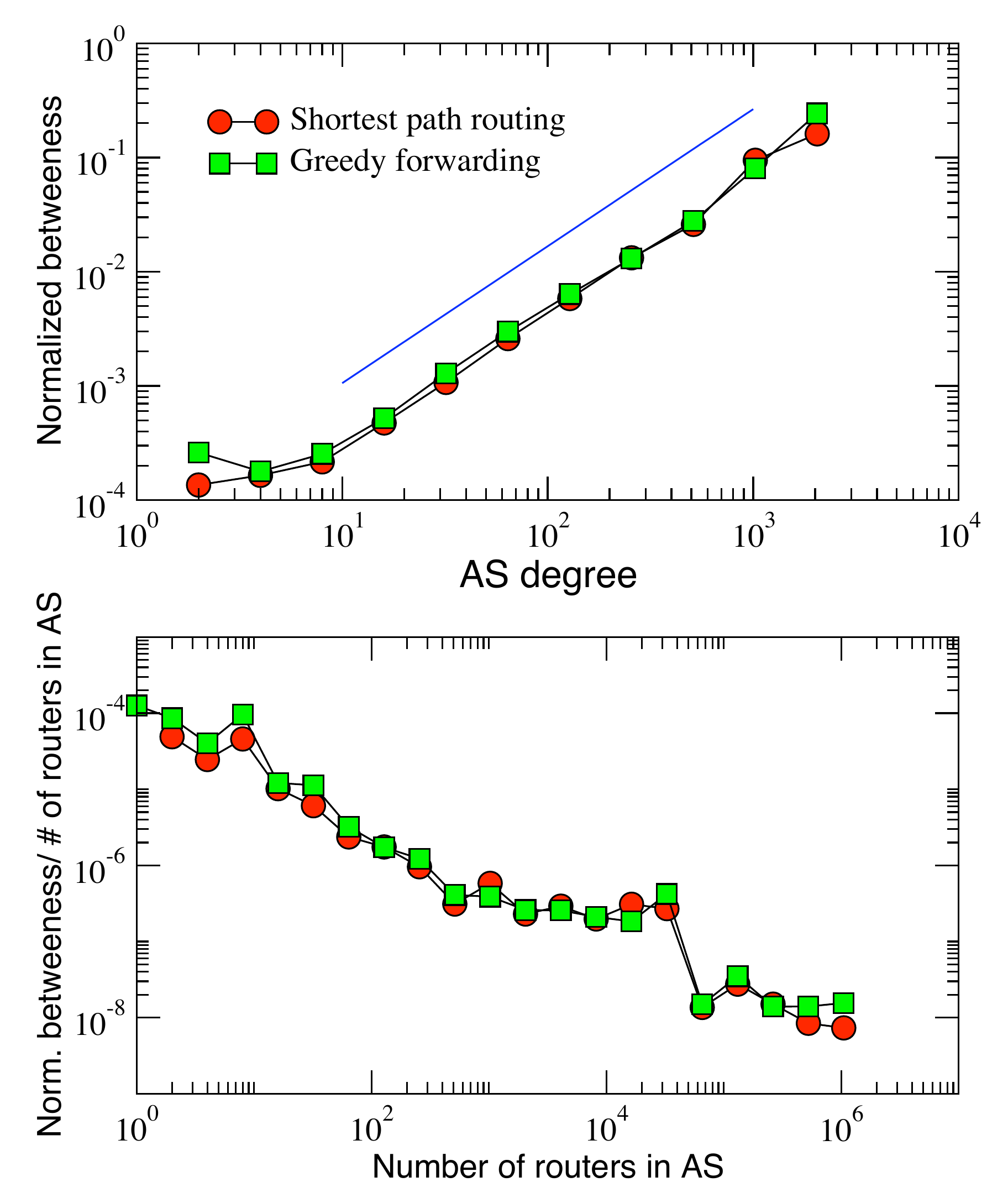}
\caption{Top: standard normalized betweenness as a function of the
AS degree for shortest path forwarding and greedy forwarding. The
solid line is a power law of exponent $1.2$. Bottom: normalized
betweenness divided by number of routers in the AS, with source and
destination ASs chosen with a probability proportional to the number
of routers in them.} \label{modelembedding9}
\end{center}
\end{figure}

In this section we measure a proxy for the amount of traffic that
ASs would have to handle under greedy forwarding.

In view of our finding that greedy forwarding follows almost always
the shortest paths, we expect that the traffic load on an AS under
greedy forwarding is essentially the same as under shortest path
forwarding. We confirm this expectation in
Fig.~\ref{modelembedding9} where we juxtapose the normalized
betweennesses corresponding to shortest path and greedy forwarding.
To compute normalized betweenness, we select a large number of
source/destination AS pairs chosen uniformly at random among all
ASs. We then find two paths for each AS pair using shortest path and
greedy forwarding. Normalized betweenness of a given AS is then the
fraction of all paths going through this AS. We observe in
Fig.~\ref{modelembedding9} that the normalized betweennesses for
shortest path and greedy forwarding are almost identical as
expected. We also observe in the top plot, that in agreement with
the previous studies on this subject, e.g.~\cite{Barthelemy04}, the
normalized betweenness grows as a power law of the AS degree. This
observation may create an impression that high-degree ASs may suffer
from traffic congestion problems. However, this impression is wrong
not only because of the results in~\cite{GkaMiSa03}, but also
because of the following considerations.

In the real Internet, ASs are not singular nodes but differently
sized networks composed of (many) routers. The size of an AS,
measured by the number of routers in it, is roughly proportional to
the AS degree~\cite{TaDoGoJaWiSh01,HuDh10}. ASs of different size
generate and consume different volumes of traffic. Also, a larger AS
can handle larger transit traffic volumes without being congested.
These two observations suggest the following modifications to the
top plot in Fig.~\ref{modelembedding9}. First, we model traffic with
the more realistic assumption that the amount of traffic an AS
generates or consumes is proportional to the AS size. That is,
instead of choosing source and destination AS pairs at random, we
chose each AS with a probability proportional to the number of
routers in the AS using the data from~\cite{HuDh10}. Second, we
divide the normalized betweenness value for each AS by the number of
routers in the AS, thus estimating the per-router traffic load. The
result shown in the bottom plot of Fig.~\ref{modelembedding9}
demonstrates that the important large ASs are, in fact, less prone
to congestion problems.

\section{Dealing with new-coming AS's}
\label{appendixD}

In this section we show that if the existing ASs keep their
hyperbolic coordinates fixed, while the ASs joining the Internet
anew over years compute their coordinates in a localizing manner,
i.e., using the LMH kernel~(\ref{loglike-local}), then the
performance of greedy forwarding does not significantly degrade,
even at long time scales.

\begin{figure}
\begin{center}
\includegraphics[width=3.5in]{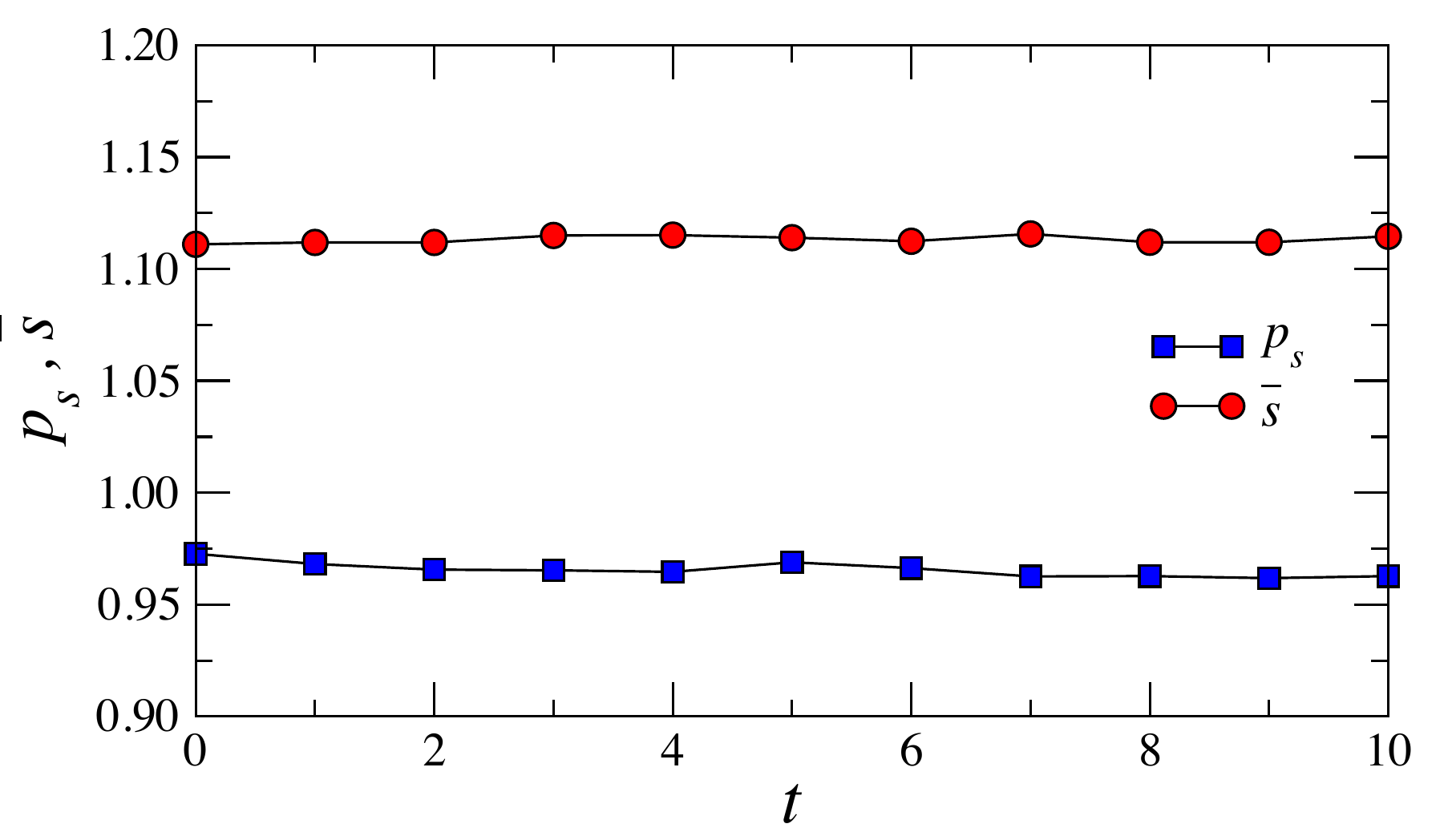}
\caption{Average success ratio $p_{s}$ and stretch
$\bar{s}$ for greedy forwarding in the AS Internet growing from
January 2007 ($t=0$) to June 2009 ($t=10$) with linear time steps
($\Delta t = 1$) of three months, e.g., $t=1$ corresponds to April
2007.\label{fig:topo_evol}}
\end{center}
\end{figure}

To demonstrate this we perform the following experiment. We replay
the AS Internet growth from January 2007 to June 2009 similar
to~\cite{DhDo08-short}. Specifically, we obtain $11$ lists of ASs observed
in the Internet at different dates as described in~\cite{DhDo08-short}.
The AS lists are linearly spaced in time with the interval of three
months: time $t=0$ corresponds to January 2007, $t=1$ is April 2007,
and so on until $t=10$, June 2009. We denote the obtained AS lists
by $A_t$. The number of ASs in $A_0$ is $17258$, while the numbers
of new ASs in $A_{t'}$ with $t'=1,2,\ldots,10$, but not in $A_0$ are
$806$, $1614$, $2389$, $3103$, $3973$, $4794$, $5434$, $5843$,
$6207$, and $6426$. We then take our Archipelago AS
topology~\cite{ClHy09-short} of June 2009, and for each $t=0,1,\ldots,10$
we remove from it all ASs and their adjacent links that are not in
$A_t$, thus obtaining a time series of historical AS topologies
$G_t$. We then embed $G_0$ using the SMH kernel~(\ref{loglike}), but
for each subsequent embedding of $G_{t'}$ with $t'>0$, we keep the
hyperbolic coordinates of ASs in $G_{t}$ with $t<t'$ fixed, and
compute coordinates for the new ASs using the LMH
kernel~(\ref{loglike-local}). That is, once an AS appears at some
time $t\geq0$ and gets its coordinates computed, using either the
SMH, $t=0$, or LMH, $t>0$, computations, the AS then never changes
its coordinates for the rest of the observation period. In
Fig.~\ref{fig:topo_evol} we show the average success ratio $p_s$ and
stretch $\bar{s}$ for greedy forwarding in $G_t$.

Remarkably, we observe only minor variations of success ratio and
stretch over more than $2.5$ years of rapid Internet growth. The
success ratio does decrease, but by less than $1\%$. We thus
conclude that greedy forwarding using our hyperbolic AS map is quite
robust with respect to Internet historical growth. Existing ASs do
not have to recompute their hyperbolic coordinates when new ASs join
the Internet. Recomputations of all AS coordinates may be executed
to improve the greedy forwarding performance, but the time scale for
such recomputations exceeds the time scale of Internet historical
evolution, i.e., years, thus exceeding by orders of magnitude the
time scale of transient dynamics of failing AS links and nodes,
i.e., seconds or minutes. That is why the existing AS coordinates
are essentially static, and can stay the same for years.

\section{Sensitivity to missing links}
\label{appendixE}

It is widely known that the existing measurements of the Internet
topology miss a number of AS
links~\cite{LaByCroXie03,DiKrFo06,OlPe10}. However, in view of the
robustness of greedy forwarding performance with respect to link
removals, one could expect that its performance would be robust with
respect to missing links as well. Furthermore, if our hyperbolic map
is used in practice, then greedy forwarding will see and use those
links that we do not see. Therefore one can intuitively expect that,
in this case, the efficiency of greedy forwarding will be actually
higher than we report in this paper, simply because these links
that we miss but greedy forwarding would not miss, would provide
additional shortcuts between potentially remote ASs. If so, the
routing results presented in this paper should be considered as
lower bounds.

To confirm this intuition, we perform the following experiment. It
is known that the majority of missing links in the Internet are
peer-to-peer links among provider ASs of moderate
size~\cite{DiKrFo06,OlPe10}. To emulate the missing link issue, we
thus remove a fraction (ranging from $0\%$ to $30\%$) of links among
nodes with degree above a certain threshold ($k=5$) from our AS
graph. We then map these graphs with different numbers of emulated
missing links to $\mathbb{H}^2$ as described in
Section~\ref{sec:AS-embedding} to find hyperbolic coordinates for
each AS. Using these maps with missing information, we then consider
two different greedy forwarding scenarios for each map:
\begin{enumerate}
\item
In the first scenario, we navigate an AS graph mapped with a
fraction of links removed, and compute the success ratio of greedy
forwarding in the graph. This scenario tries to mimic the missing
links issue directly. We have incomplete topology measurements of
the real Internet, but we have no other option as to use these
measurements to map the Internet to its hyperbolic space, and study
navigability with this map, which we know miss some information.
\item
In the second scenario, we use the hyperbolic map obtained with
missing links, but we then add back those removed links, and
navigate the complete graph. This scenario is motivated by the
observation that even though our map is constructed with some links
missing, these missing links will still be used by ASs attached to
them to forward information if this map is used in practice.
\end{enumerate}

\begin{figure}
\begin{center}
\includegraphics[width=3.5in]{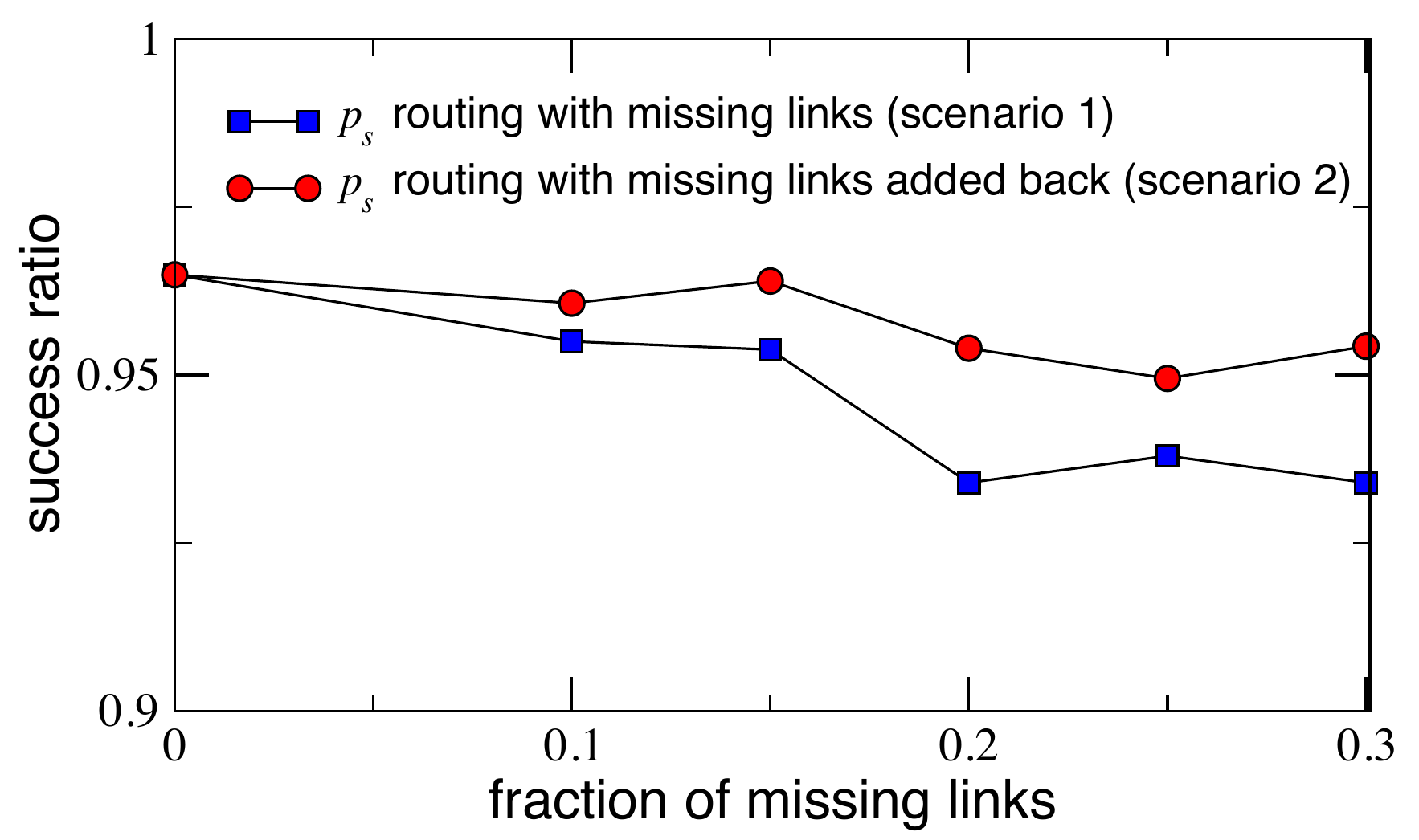}
\caption{Success ratio as a function of the fraction of
removed links among nodes of degree above $5$ for the two scenarios
described in the text. Note that $30\%$ of missing links in this
subgraph corresponds to $14\%$ of the total number of links in the
network. } \label{modelembedding10}
\end{center}
\end{figure}

The results of these two scenarios are shown in
Fig.~\ref{modelembedding10}. As intuitively expected, our mapping is
quite robust with respect to missing links: the success ratio
decreases by less than $5\%$ even if up to $14\%$ of links are
removed from the topology before we map it. Also as expected, the
missing links, when added back, increase the success ratio. That is,
even though the map has been constructed using partial information,
navigability improves when missing links are considered. These
results confirm that the routing results reported in this paper
are in reality lower bounds for the success ratio that can be
achieved if our map is used in practice. In fact, one may somewhat
paradoxically expect that the more links are missed in the measured
Internet topology we used for mapping, the better the success ratio
would be in practice, since according to
Fig.~\ref{modelembedding10}, the success ratio improvement due to
re-adding of removed links tends to increase with the number of
removed links.

\section{Comments on AS-level routing}

Our approach belongs to a wide class of approaches proposing to
reduce routing granularity to the level of Autonomous
Systems~\cite{nimrod,encaps,islay,atoms,GuKoKi04,SuKaEe05,CaCo06,KrKc07,OlLa07,MaWa07,ZhaZha09,lisp,ShGu10}.
The key difference between ours and the existing approaches in this
class is that the latter require some form of routing on the dynamic
AS graph. As soon as the AS topology changes, new AS routes must be
recomputed, so that routing communication overhead is unavoidable in
this case. In our case such recomputations are not needed since as
we have shown, the efficiency of greedy forwarding
sustains in presence of failing AS nodes and links, even though ASs
do not exchange any information about topology modifications, and do
not change their hyperbolic coordinates, i.e., even though they do
not incur any communication overhead. A bulk of routing overhead in
the Internet today is due to traffic engineering and multihoming in
the first place~\cite{huston01-01,huston06}. How the AS-level
routing class of approaches helps to deal with and reduce this
overhead is discussed in the literature cited above.

\begin{acknowledgments}
 We thank M.~Newman and M.~\'Angeles Serrano for many useful suggestions and
discussions, M.~\'Angeles Serrano for suggesting the analogy with
gravitation, A.~Aranovich for help with Fig.~\ref{fig:0}, and
Y.~Hyun, B.~Huffaker, and A.~Dhamdhere for help with the data.
M.~B.\ acknowledges support from DGES grant
No.~FIS2007-66485-C02-02, Generalitat de Catalunya grant
No.~2009SGR838, and NSF CNS-0964236. D.~K.\ acknowledges support
from NSF CNS-0722070 and CNS-0964236, DHS N66001-08-C-2029, and
Cisco Systems.
\end{acknowledgments}

\end{document}